\documentclass[aps,pre,reprint,a4paper,nofootinbib, superscriptaddress]{revtex4-1}
\usepackage{cmap}
\usepackage[T1]{fontenc}
\usepackage{newtxtext}
\usepackage[slantedGreek]{newtxmath}

\pdfpageattr{/Group <</S /Transparency /I true /CS /DeviceRGB>>}

\usepackage[version=3]{mhchem}
\usepackage{siunitx}
\DeclareSIUnit\Molar{\textsc{m}}
\usepackage{mleftright}
\usepackage[pdftex,bookmarks, pdffitwindow=false, pdfstartview=FitH, pdfdisplaydoctitle, colorlinks, plainpages=false,pdftitle={Direct observation and rational design of nucleation behavior in addressable self-assembly},pdfauthor={Martin Sajfutdinow, William M. Jacobs, Aleks Reinhardt, Christoph Schneider, David M. Smith}, pdfpagelabels, hypertexnames, citecolor={blue},linkcolor={red}, urlcolor={violet!50!black}, pdflang={en}, hyperfootnotes=false, breaklinks]{hyperref}
\usepackage{graphicx}
\usepackage{amsmath,amssymb}
\usepackage{xtab, booktabs}

\usepackage{enumitem}

\usepackage{geometry}
\usepackage[stretch=17,shrink=17,step=1]{microtype}

\newcommand{\refSub}[2]{\hyperref[#2]{\ref{#2}#1}}
\newcommand{\figref}[1]{Fig.~\ref{#1}}
\newcommand{\figrefsub}[2]{Fig.~\refSub{#2}{#1}}
\newcommand{\secref}[1]{Sec.~\ref{#1}}
\newcommand{\SIseczref}[1]{Sec.~SI-\ref{#1}}
\newcommand{\SIfigzref}[1]{Fig.~\ref{#1}}

\newcommand{\zref}[1]{\ref{#1}}

\usepackage[version=3]{mhchem}
\usepackage{siunitx}
\DeclareSIUnit\Molar{\textsc{m}}

\newcommand{\der}{\ensuremath{\mathrm{d}}}
\newcommand{\Rh}{\ensuremath{R_{\text{h}}}}
\newcommand{\TM}{\ensuremath{T_{\text{m}}}}
\newcommand{\rhom}{\rho_{\text{m}}}

\newcommand{\rhoT}{\rho_{\text{T}}}

\usepackage{xcolor}
\colorlet{colourAll}{blue!90!black}
\colorlet{colourEdge}{red!90!black}
\colorlet{colourFace}{green!75!black}
\colorlet{colourNo}{black}
\colorlet{colourScaff}{orange!50!black}
\definecolor{colourHalfFace}{RGB}{0, 127, 255}

\makeatletter
\def\@bibdataout@aps{
 \immediate\write\@bibdataout{
 @CONTROL{
   apsrev41Control, author="08",editor="1",pages="0",title="0",year="1"
 }}
 \if@filesw
  \immediate\write\@auxout{\string\citation{apsrev41Control}}
 \fi
}
\makeatother

\begin{document}

\title{Direct observation and rational design of nucleation behavior\\in addressable self-assembly}
\author{Martin Sajfutdinow}
\affiliation{Fraunhofer Institute for Cell Therapy and Immunology IZI, Department of Diagnostics, Perlickstra\ss{}e 1, 04103 Leipzig, Germany}
\affiliation{Faculty of Chemistry and Mineralogy, Leipzig University, Johannisallee 29, 04103 Leipzig, Germany}

\author{William M. Jacobs}
\affiliation{Department of Chemistry and Chemical Biology, Harvard University, 12 Oxford Street, Cambridge, Massachusetts 02138, United States}

\author{Aleks Reinhardt}
\affiliation{Department of Chemistry, University of Cambridge, Lensfield Road, Cambridge, CB2 1EW, United Kingdom}

\author{Christoph Schneider}
\affiliation{Fraunhofer Institute for Cell Therapy and Immunology IZI, Department of Diagnostics, Perlickstra\ss{}e 1, 04103 Leipzig, Germany}

\author{David M. Smith}
\affiliation{Fraunhofer Institute for Cell Therapy and Immunology IZI, Department of Diagnostics, Perlickstra\ss{}e 1, 04103 Leipzig, Germany}

\begin{abstract}
  In order to optimize a self-assembly reaction, it is essential to understand the factors that govern its pathway.
  Here, we examine the influence of nucleation pathways in a model system for addressable, multicomponent self-assembly based on a prototypical `DNA-brick' structure.
  By combining temperature-dependent dynamic light scattering and atomic force microscopy with coarse-grained simulations, we show how subtle changes in the nucleation pathway profoundly affect the yield of the correctly formed structures.
  In particular, we can increase the range of conditions over which self-assembly occurs by utilizing stable multi-subunit clusters that lower the nucleation barrier for assembling subunits in the interior of the structure.
  Consequently, modifying only a small portion of a structure is sufficient to optimize its assembly.
  Due to the generality of our coarse-grained model and the excellent agreement that we find with our experimental results, the design principles reported here are likely to apply generically to addressable, multicomponent self-assembly.
\end{abstract}

\maketitle

Increasingly complex structures can now be created by self-assembly~\cite{Whitelam2015, Frenkel2015}, from nanostructures with tailored physicochemical properties, such as photonic crystals~\cite{Zhang2009, *Sowade2016}, to quasicrystals~\cite{Urgel2016,*Reinhardt2017,*Damasceno2017}.
In the limit where every subunit in a target structure is unique and bonds strongly with specific partners, such self-assembled structures are said to be `addressable'.
Thus far, this degree of specificity has been demonstrated most impressively by experiments on `DNA bricks'~\cite{Ke2012,*Ong2017}, in which portions of single-stranded DNA molecules are designed to hybridize uniquely with complementary sequences on strands that occupy neighboring positions in the target structure.
Modular nanostructures comprising thousands of distinct strands can be formed in this way, and because the location of each molecule in the target structure is precisely known, these structures can be functionalized at a nanometer length scale.

In addition to providing control over the geometry of the target structure, the use of addressable building blocks makes it possible to exert greater control over the \textit{mechanism} of self-assembly~\cite{Jacobs2016}.
Because each interaction between subunits can be individually tuned, addressable structures provide a useful platform for exploring the determinants of self-assembly pathways more generally~\cite{Cademartiri2015}.
Considerable progress has been made in this direction using computer simulations~\cite{Schulman2010, *Zenk2014, Reinhardt2014, Zeravcic2014, Madge2015, *Madge2017, Reinhardt2016, Reinhardt2016b,WaymentSteele2017, Wales2017, Fonseca2017} and statistical mechanics~\cite{Jacobs2015, Jacobs2015b,Jacobs2015c} to study coarse-grained models of addressable systems.
In particular, coarse-grained modeling has predicted that nucleation barriers\footnote{The term `nucleation' in the context of DNA self-assembly is occasionally used to refer to the initial thermodynamically disfavored formation of a few base pairs of a double strand, which is then followed by zipping~\cite{Pinheiro2012}. We use the term in the more traditional sense to mean the formation of a small portion of the target structure, which leads to structure assembly.} are likely to play a particularly important role in addressable self-assembly, since in their absence, the large number of building blocks with similar bonding strengths can instead lead to widespread kinetic trapping and aggregation~\cite{Reinhardt2014,Jacobs2015, *Jacobs2015b}.
These models have further shown that addressable systems often have highly non-classical nucleation barriers and well-defined critical nuclei~\cite{Jacobs2015, Jacobs2015b, Reinhardt2016b}.
However, the microscopic nature of a self-assembly process is challenging to study experimentally.
While it is possible to characterize structures by stopping the reaction at a specific point along an annealing ramp~\cite{Sobczak2012,Majikes2017} for subsequent imaging~\cite{Song2012,Kato2009}, such approaches cannot be performed \textit{in situ} and may thus perturb the self-assembly process.
Furthermore, any assembled structures must first be isolated before carrying out more detailed analyses, for example by using next-generation sequencing to examine defects in DNA nanostructures~\cite{Myhrvold2017}.
On the other hand, established \textit{in situ} methods can provide information on the kinetics of self-assembly, but only by probing the interactions between pairs of subunits~\cite{Wei2013b,Jiang2017,Pinheiro2012,Sobczak2012,Wei2014}.
As a result, these interactions must then be extrapolated to describe the assembly of the complete structure.

\begin{figure*}[tbp]
  \includegraphics{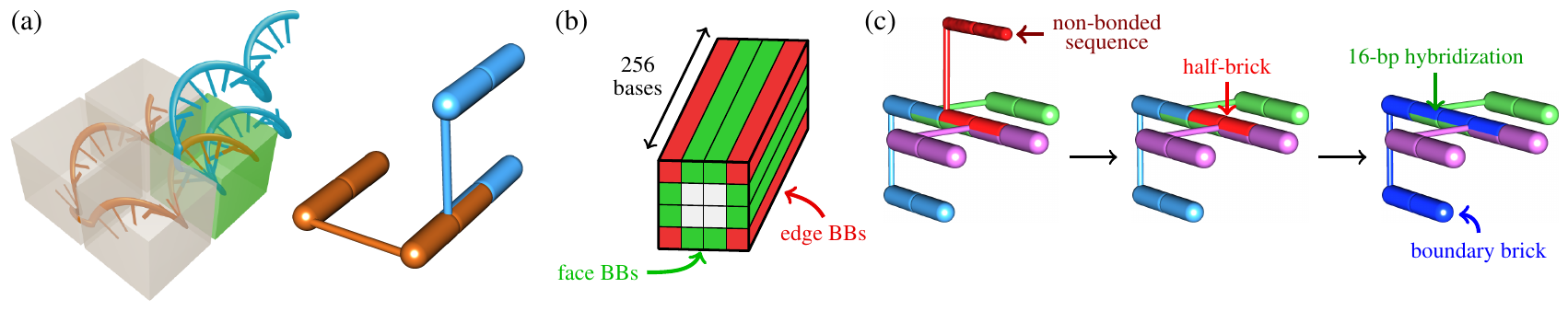}
\caption{
(a) The bonding pattern between two DNA bricks in a strand and a cylindrical representation. Each molecule is partitioned into four domains (indicated by boxes), while the neighboring bricks are bonded through one pair of domains only (green box).  The cylindrical representation shows the same 8-bp hybridization.
(b) A schematic illustration of our target structure, highlighting the locations of the `edge' and `face' boundary bricks.
(c) A schematic of the boundary brick set-up: the non-interacting DNA sequence at one of the outer surfaces of the target structure is removed and the remainder of that brick is fused with the adjacent brick, resulting in the formation of a 48-nt boundary brick.
}\label{fig-schematic-bonding}
\end{figure*}

\begin{figure*}[t!]
\includegraphics{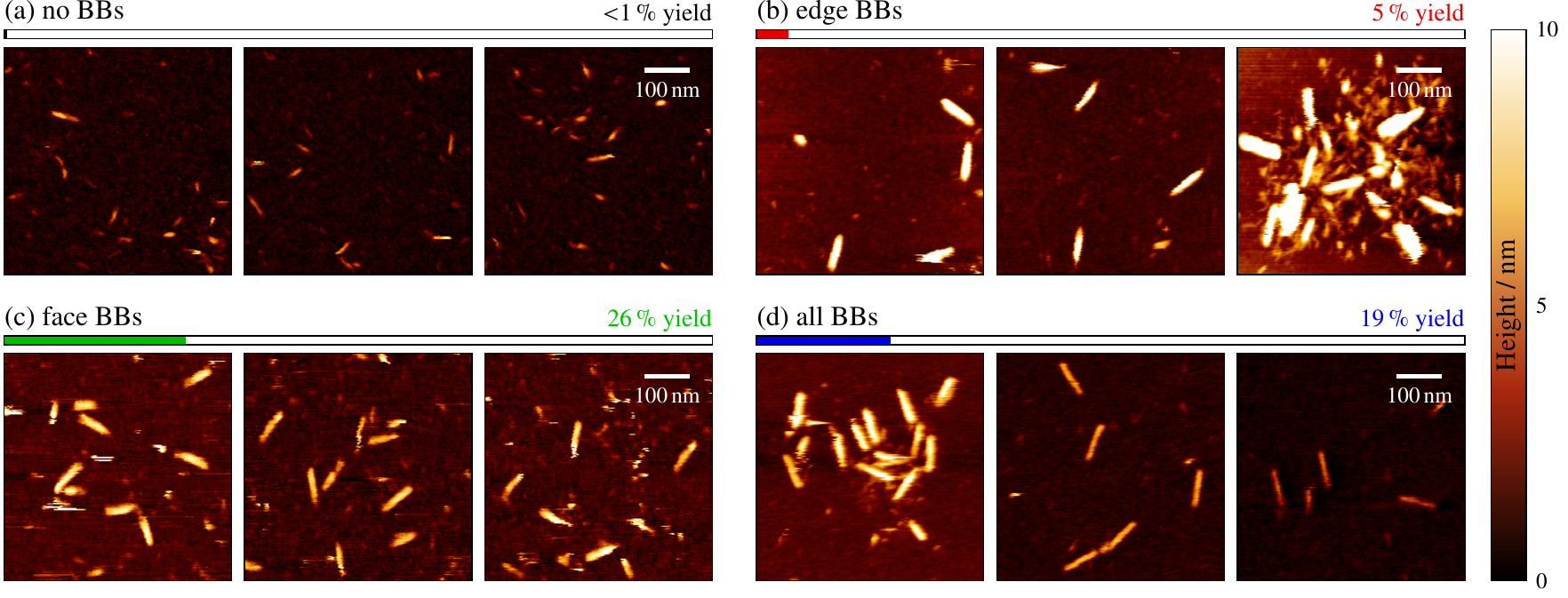}
\vskip-1.75ex
\caption{AFM images taken at the conclusion of a \SI{66}{\hour} annealing protocol for each of the four \SI{86}{\nano\metre}-long cuboid designs.
No purification was performed on these samples so that incompletely assembled structures can clearly be seen. Typical yields, relative to the total quantity of DNA strands in solution, were estimated via gel electrophoresis and are shown for each system. For the edge-BB system in particular, multi-structure aggregates (shown in the right-most panel) are commonly observed.}\label{fig-AFM}
\vskip-0.75ex
\end{figure*}

In this work, we demonstrate that dynamic light scattering (DLS) can be used to track the collective assembly of addressable structures in greater detail.
Unlike alternative \textit{in situ} techniques, DLS provides a sensitive means of probing the complete distribution of multi-strand cluster sizes throughout the course of the annealing protocol.
Consequently, by applying DLS to DNA-brick self-assembly and validating these results using atomic force microscopy (AFM), we are able to analyze the nucleation process as a function of temperature and assembly time.
Combining these results with extensive simulations, we show that it is possible to control the nucleation behavior rationally, with dramatic consequences for the yield of assembled structures.
In particular, we demonstrate that the self-assembly mechanism can be optimized by altering the connections among a specific subset of subunits, which modifies the free-energy barrier for structure nucleation.
The simplicity of our coarse-grained model suggests that these design principles are transferable to any multicomponent system where the interactions between subunits can be programmed.

\section*{Results}

\subsection*{Minor changes in nanostructure design strongly affect the yield and quality of self-assembly}

As a model system, we examined the self-assembly of a 16-helix DNA cuboid.
Following the canonical `DNA-brick' design~\cite{Ke2012}, the fundamental building blocks of this structure are 32-nucleotide (nt) `scaffold' bricks.
Each brick comprises four 8-base-pair (bp) domains that hybridize to connect adjacent helices (\figrefsub{fig-schematic-bonding}{a}).
The cross-section was chosen to ensure that bricks on opposite sides of the structure do not interact directly (\figrefsub{fig-schematic-bonding}{b}), while the high aspect ratio (4 helices $\times$ 4 helices $\times$ 256 bases) facilitates the identification of well-formed structures via atomic force microscopy imaging.

To study the factors affecting the self-assembly yield, we designed variants of this cuboid by increasing the lengths of a small number of complementary domains.
This was achieved by varying the numbers and types of 48-nt `boundary bricks' (BBs) at the exterior surfaces of the structure (\figrefsub{fig-schematic-bonding}{c}; see also \SIfigzref{fig-BB-illustrations}).
In addition to a cuboid composed entirely of scaffold bricks (`no BBs'), where the 16-bp half-bricks at the exterior of the structure were left unconjugated, we designed variants with boundary bricks forming the corner helices (`edge BBs'), connecting pairs of helices on the faces of the cuboid (`face BBs') or both (`all BBs').
All variants of the cuboid structure self-assembled to some degree over the course of a \SI{66}{\hour} linear annealing ramp (see \SIseczref{subsec-annealing}).
However, AFM imaging (see \SIseczref{subsec-afm-protocols}) revealed striking differences in the quality of the assembled structures (\figref{fig-AFM}).
The all-BB, face-BB and edge-BB designs resulted in the assembly of many copies of structures with the expected aspect ratio, while designs without boundary bricks yielded a negligible number of such structures (see \SIseczref{subsec-gel-electrophoresis} and \SIfigzref{fig-gels}).

\subsection*{Tracking structure assembly via DLS}

To obtain information on the self-assembly process, we used DLS to probe the size of structures as a function of temperature during the annealing ramp.
These measurements provide insight into the growth of clusters of hybridized strands without requiring the introduction of intercalating dyes or other additives that might alter structure assembly.
Because sub-micron-sized particles scatter visible light in the Rayleigh limit, where the scattering intensity scales as the sixth power of the particle size, DLS is also highly sensitive to small populations of large clusters.
These features of DLS therefore allowed us to detect the initial formation of the target cuboids during the annealing protocol without perturbing the assembly process.

At each temperature step, we obtained the auto-correlation function from a time series of light-scattering intensity measurements in order to extract a distribution of decay rates.
Given the low concentration of macromolecules in our experiments ($\sim\!\SI{0.2}{\percent}$ by volume), we assumed that the free diffusion of particles in the suspension was not affected by hydrodynamic interactions, so that the decay rates could be related to the translational diffusion coefficients of independent multi-strand clusters~\cite{Berne1976}.
For ease of interpretation, we present these distributions in terms of the hydrodynamic radius \Rh{} of a spherical particle with an equivalent diffusion coefficient.
Since determining the decay rate distribution from the auto-correlation function requires additional assumptions on the smoothness of the cluster-size distribution, we used multiple regularization methods to verify the robustness of our results (\SIfigzref{fig-DLS-regularisation}; see \SIseczref{subsec-scattering}).

\begin{figure}[t!]
\includegraphics{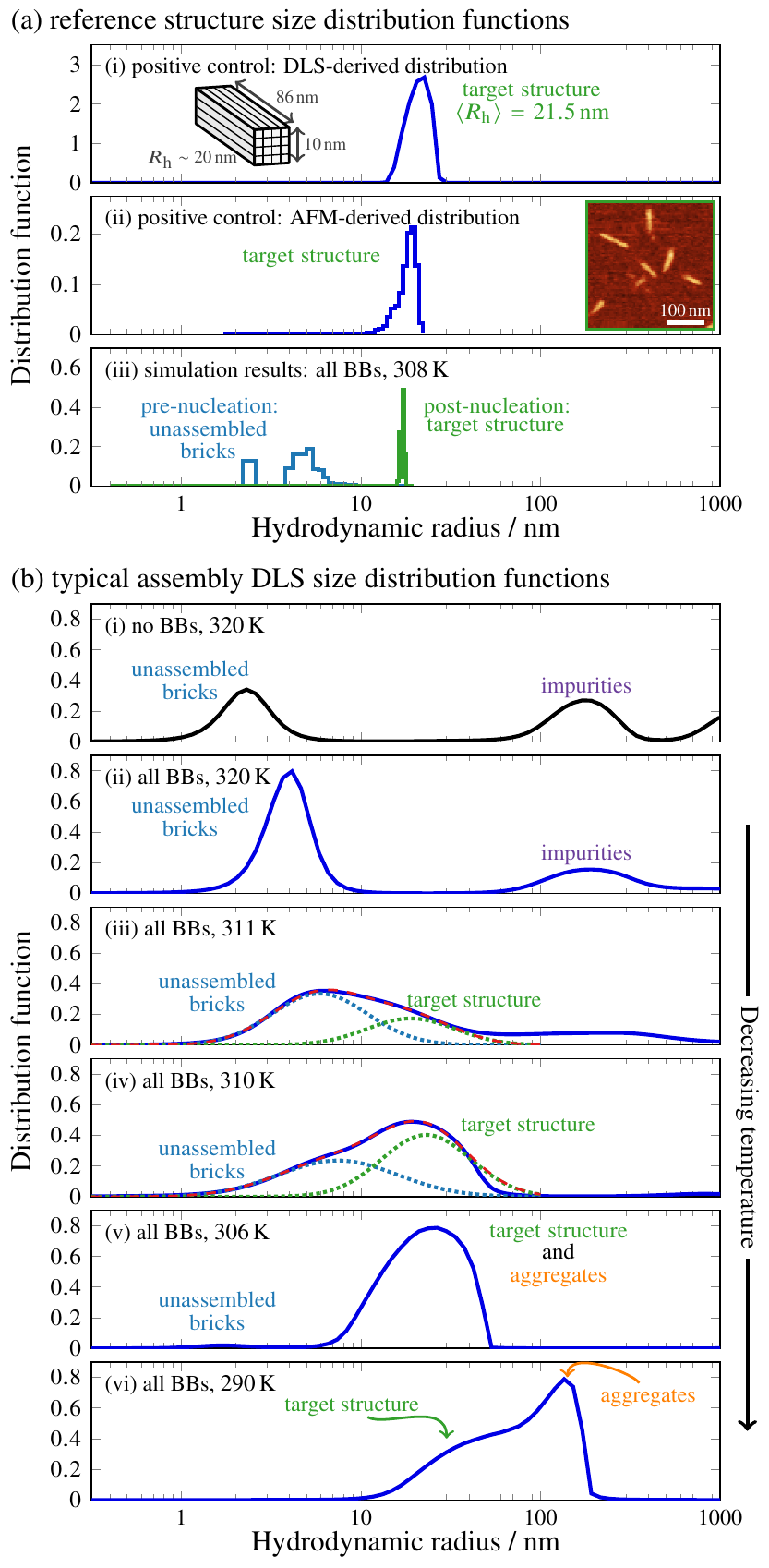}
\caption{(a) Reference intensity-weighted size distribution functions for a purified all-BB sample, determined by DLS (panel~i) and calculated from AFM images (panel~ii) and Monte Carlo simulations (panel~iii).  Insets show the dimensions of the target structure and a representative AFM image.  In panel~iii, the blue curves indicate the metastable cluster-size distribution prior to nucleation, while the green curves show the equilibrated distribution after the target structure has assembled.  The gap between the monomer and dimers in the pre-nucleation distribution is an artifact of the lattice simulations.  (b) Representative intensity-weighted distribution functions at decreasing temperatures.  At high temperatures (panels~i and ii), the no-BB distribution indicates unhybridized scaffold strands, while the all-BB distribution is dominated by small clusters of BBs, as seen in the pre-nucleation Monte Carlo simulations.  At intermediate temperatures (panels~iii and iv), the distribution can be fitted to a sum of two Gaussians (red dashes), which correspond to the unassembled (blue dots) and target-structure (green dots) populations, respectively.  At lower temperatures (panels~v and vi), a small population of larger aggregates skews the intensity-weighted distribution, but the contribution from the target structure can still be seen in panel~vi.}
\label{fig-DLS-validation}
\end{figure}

We first determined the reference cluster-size distribution for a purified sample of assembled all-BB cuboids (\figrefsub{fig-DLS-validation}{a,i}).
This distribution is peaked at a hydrodynamic radius of \SI{21.5}{\nano\metre}, which matches the expected size of a fully assembled cuboid ($\Rh\!\sim$\SI{20}{\nano\metre}; see Methods).
This distribution also agrees with the ideal distribution calculated from AFM images of purified all-BB cuboids (\figrefsub{fig-DLS-validation}{a,ii}), in which all imaged particles were treated as rigid cylinders (see Methods).
The broadening of the reference distribution relative to this ideal distribution is likely due to the effects of particle anisotropy on light scattering, which we have not attempted to account for here.

We next used lattice Monte Carlo simulations of an established coarse-grained model~\cite{Reinhardt2014} to calculate ideal cluster-distribution functions of the all-BB system, equilibrated both before and after nucleation of the target structure (\figrefsub{fig-DLS-validation}{a,iii}; see Methods).
Consistent with prior simulations~\cite{Reinhardt2014,Jacobs2015,Jacobs2015b}, we found that intermediate cluster sizes, with \Rh{} between 8 and \SI{15}{\nano\metre}, are unstable.
Consequently, the size distribution is either peaked near \SI{5}{\nano\metre}, corresponding to small clusters of primarily boundary bricks, or \SI{18}{\nano\metre}, corresponding to a mostly complete target structure.
Because these simulations consider a single copy of the target structure, the system can only be in one state at a time; however, in a larger system with many copies of each brick, the assembly of a fraction of all structures would result in a bimodal cluster-size distribution.
The simulation results therefore suggest that the \Rh{} distribution can be used to resolve the target structure during an assembly experiment.
We note that the discretization of small cluster sizes in the unassembled population is an artifact of the lattice model and is not expected to be seen in experiments.

Typical size-distribution functions determined by DLS similarly show that \Rh{} is a suitable order parameter for identifying complete structures (\figrefsub{fig-DLS-validation}{b}).
At high temperatures (\figrefsub{fig-DLS-validation}{b,i-ii}), before nucleation occurs, we observed a single peak (ignoring high-molecular-weight impurities) corresponding to individual strands and small clusters.
In particular, in the no-BB system, the peak matches the expected size of a flexible 32-nt strand, $\Rh\!\sim$\SI{2.7}{\nano\metre}.
Then, upon decreasing the temperature, a new population suddenly appeared at $\Rh\!\sim$\SI{20}{\nano\metre}.
As expected from our simulation results, the cluster-size distributions at these intermediate temperatures are well described by bimodal fits to a linear combination of Gaussian functions (\figrefsub{fig-DLS-validation}{b,iii-iv}).
In particular, the means of the Gaussian fits coincide with the reference unassembled and target-structure distributions; however, the fitted populations are considerably broader than the reference distributions.
This is likely a consequence of the conservative regularization method used in the analysis of the autocorrelation data, which tends to smooth the resulting distributions, as well as heterogeneity due to incomplete assembly.
To confirm our interpretation of the bimodal cluster-size distributions, we discuss a complementary validation strategy based on an analysis of AFM images below and in the Supplementary Material.

At lower temperatures (\figrefsub{fig-DLS-validation}{b,v-vi}), particles with effective hydrodynamic radii larger than $\sim$\SI{40}{\nano\metre} begin to contribute to the distribution.
This shift toward larger \Rh{} is likely due to the formation of aggregates of fully or partially assembled structures.
However, we emphasize that because of the sixth-power dependence of the light scattering intensity on the particle size, only a small fraction of aggregated structures is needed to skew the cluster-size distribution substantially.
For the same reason, the large-\Rh{} impurities present at higher temperatures (\figrefsub{fig-DLS-validation}{b,i,ii}) are extremely rare.
Nevertheless, despite the tendency of the structure and aggregate peaks to merge due to our conservative choice of regularization method, the population of target structures can still be identified from the shoulder of the cluster-size distribution at \SI{290}{\kelvin} (\figrefsub{fig-DLS-validation}{b,vi}).

\subsection*{Evidence for nucleation and growth}

\begin{figure}[t!]
  \includegraphics{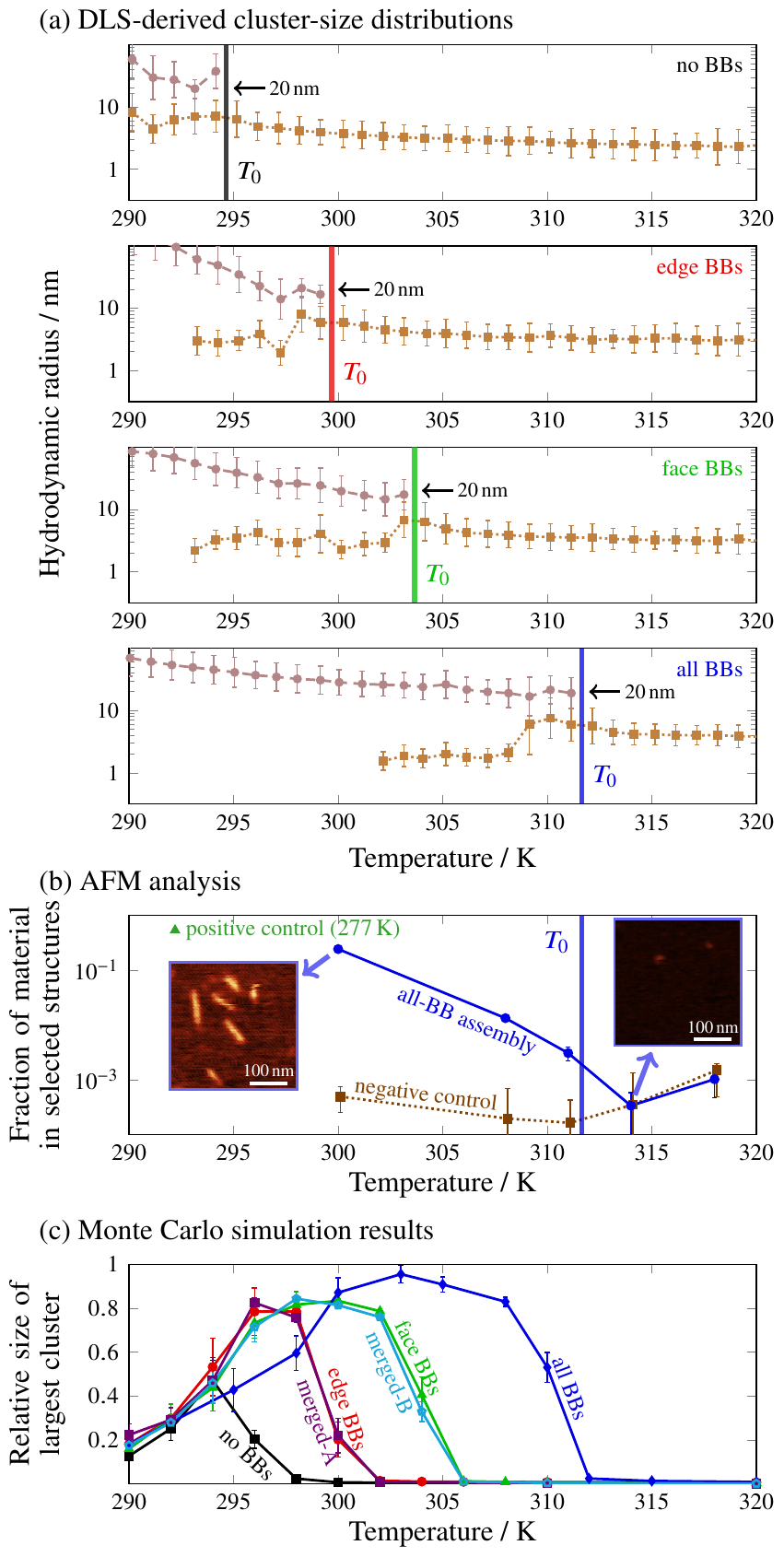}
  \caption{(a) Unimodal and bimodal cluster-size distributions determined via DLS following a \SI{15.2}{\hour} annealing protocol.  Points and error bars show the mean hydrodynamic radii and standard deviations, respectively, of the Gaussian fits to the unassembled and target structure populations (cf.~\figrefsub{fig-DLS-validation}{b,iii-iv}) at each temperature.  For each system, $T_0$ indicates the temperature at which the higher-\Rh{} population first appears during the annealing protocol.  (b)  The volume fraction comprising target cuboids, relative to the total volume of material, determined from AFM analysis of the all-BB system.  All samples were rapidly quenched for imaging from the indicated temperatures.  The target-structure volume fraction of the positive-control all-BB sample (green triangle) and non-hybridizing oligonucleotides (negative control, dotted line) were calculated in the same way, and the error bars show the estimated standard error based on Poisson statistics.  Insets show representative portions of the raw AFM images.  (c) The size of the largest correctly bonded cluster from Monte Carlo simulations. Each data point corresponds to the average of ten independent simulations in the long time limit, once nucleation has occurred, with error bars showing the standard deviation. The `merged' curves refer to fixed edge dimers, as discussed in the text and \SIfigzref{fig-BB-illustrations}.}
  \label{fig-cluster-size-analysis}
\end{figure}

For each structure variant, we determined both the cluster-size distribution via DLS and the extent of subunit hybridization via fluorescence measurements (see \SIseczref{subsec-fluorescence}) as a function of temperature over the course of a \SI{15.2}{\hour} linear annealing protocol.
We observed prominent peaks in the fluorescence response of systems containing BBs at high temperatures ($>\SI{330}{\kelvin}$).
This behavior could be attributed to the formation of stable high-temperature dimers, in which the elongated boundary bricks (\figrefsub{fig-schematic-bonding}{c}) stably hybridize to other bricks in continuous 16- or 24-bp domains (\SIfigzref{fig-hybrid}).
However, in DLS experiments, we did not observe any substantial change in the overall scattering intensity at temperatures above \SI{315}{\kelvin} (\SIfigzref{fig-cuboid-static}), implying that the assembly of complete structures does not take place at these temperatures.
Nevertheless, DLS did resolve differences in the unassembled populations.
At temperatures where only a single peak was present (excluding contributions from any impurities in the system), we found that the mean hydrodynamic radius $\langle\Rh\rangle$ of the no-BB system increased from $\sim\!\SI{2.5}{\nano\metre}$ to $\sim\!\SI{4}{\nano\metre}$ upon cooling, reflecting an increasing fraction of scaffold-strand dimers (\figrefsub{fig-cluster-size-analysis}{a}).
Similarly, the single-peak $\langle\Rh\rangle$ in systems with boundary bricks increased from $\gtrsim\!\SI{3}{\nano\metre}$ to $\sim\!\SI{5}{\nano\metre}$ upon cooling, consistent with the presence of larger pre-formed BB dimers.

In each system, we observed the sudden appearance of a second peak in the cluster-size distribution at a temperature $T_0$ (\figrefsub{fig-cluster-size-analysis}{a}).
This feature appeared at the same temperature in multiple annealing runs for each system, with the exception of the no-BB structure, where $T_0$ varied by $\sim\!\SI{2}{\kelvin}$ across three runs.
As in the example distributions shown in \figrefsub{fig-DLS-validation}{b,iii-iv}, the mean hydrodynamic radius of this population, determined by fitting a linear combination of Gaussian functions, coincided with the expected size of the target structure in all systems.
Because of the comparable scattering intensities of the two populations at $T_0$, we ascribed this second peak to the scattering of a relatively small number of essentially complete target structures.
The target-structure $\langle\Rh\rangle$ remained nearly constant for at least \SI{3}{\kelvin} below $T_0$ in all systems before increasing above \SI{20}{\nano\metre}, most likely due to aggregation as discussed above.
By contrast, the fluorescence response (\SIfigzref{fig-hybrid}) did not provide definitive insights into the assembly of the complete structure for any cuboid variant.

Importantly, our experiments indicate that the target structures do not grow gradually as a function of temperature.
Instead, DLS reveals that the transition from having all unassembled subunits to having some complete structures occurs discontinuously.
The unassembled population remains easily detectable over a temperature range of approximately \SI{10}{\kelvin} below $T_0$ for each structure, indicating that not all subunits are incorporated into complete structures at $T_0$.
For the no-BB, edge-BB, and face-BB systems, the mean \Rh{} of this population is comparable to the mean \Rh{} of unassembled strands above $T_0$.
By contrast, the mean \Rh{} of this population decreased in the all-BB system near \SI{308}{\kelvin}, suggesting that only scaffold strands remained unassociated with target structures or aggregates below this temperature.

To validate further our interpretation of the cluster-size distributions obtained from DLS, we performed a complementary analysis based on AFM imaging of the all-BB system at selected temperatures.
Using AFM images of quenched and immobilized samples, we estimated the fraction of the total volume of imaged particles comprising target structures.
We first determined appropriate criteria, using the areas and aspect ratios of imaged particles, for identifying correctly assembled cuboids in images of a purified sample (\SIfigzref{fig-AFM-all-BB-positive-control}).
We then applied these criteria to estimate the target-structure volume fraction as a function of the temperature from which the sample was quenched (\figrefsub{fig-cluster-size-analysis}{b}; see also \SIfigzref{fig-AFM-all-BB-images-distributions}).
Because the rapid quenching involved in the preparation of the samples likely affects the particle-size distribution and AFM does not reliably distinguish single-stranded DNA from the background, this method cannot be used to assess the volume fraction of assembled structures quantitatively.
In addition, image analysis inevitably identifies some false target structures.
However, by comparing the calculated volume fractions with a negative control of approximately 200 non-hybridizing, similar-length oligonucleotides, which accounts for sample-preparation and imaging artifacts, we verified that the target structures are indeed present at temperatures below, but not above, $T_0$.
The low estimated volume fraction (approximately \SI{0.3}{\percent}) just below $T_0$ is also consistent with the roughly equal areas of the intensity-weighted unassembled and target-structure populations seen in DLS (\figrefsub{fig-DLS-validation}{b,iii-iv}), demonstrating the sensitivity of DLS to small populations of large clusters.
This analysis therefore corroborates our primary conclusions from the DLS experiments and supports our interpretation of the ${\Rh\simeq\SI{20}{\nano\metre}}$ population.
The remainder of our study is based on DLS data, since this technique can be performed \textit{in situ} without perturbing the assembly process.

\subsection*{Comparison with coarse-grained Monte Carlo simulations}

To observe the self-assembly process in greater detail, we simulated the assembly of a coarse-grained DNA-brick model using Monte Carlo dynamics at constant temperature~\cite{Reinhardt2014}.
Previous studies~\cite{Reinhardt2014,Jacobs2015b} of this model have found that self-assembly proceeds via nucleation and growth, whereby clusters that are intermediate between unassembled strands and nearly complete target structures are thermodynamically unstable.
In particular, the nucleation step, which requires the formation of a critical multi-strand cluster, is a thermally activated rare event and thus determines the highest temperature at which self-assembly can occur.
Therefore, following an approach established for simulating structures with BBs~\cite{WaymentSteele2017}, we studied the nucleation of cuboid designs analogous to those used in our DLS experiments, using a single copy of the target structure and hybridization parameters chosen to mimic the experimental conditions (see Methods).

Remarkably, we found that for each cuboid variant, the highest temperature at which nucleation occurs in our simulations is in nearly quantitative agreement with the temperature at which the $\Rh\simeq\SI{20}{\nano\metre}$ population first appears in the DLS experiments.
This can be seen by comparing the temperature at which the average cluster size sharply increases in \figrefsub{fig-cluster-size-analysis}{c} with the corresponding $T_0$ in \figrefsub{fig-cluster-size-analysis}{a}.
It is important to note that, unlike in the experiments, all simulations were initialized from an unassembled solution with the total experimental monomer concentration at each temperature.
The simulated trajectories should thus only be compared to the initial formation of target structures near $T_0$ during the annealing ramp, after which monomer depletion must be taken into account.
In simulations initiated at lower temperatures, kinetic trapping arising from subunit misbonding tends to inhibit structure nucleation, as evidenced by the decreased average cluster sizes at temperatures below $\sim\!\SI{295}{\kelvin}$ (\figrefsub{fig-cluster-size-analysis}{c}).
In contrast with the variations in nucleation behavior, the effects of misbonding are essentially independent of the structure design in our simulations.

\subsection*{Pre-formed clusters modify nucleation barriers}

\begin{figure}[t!]
\includegraphics{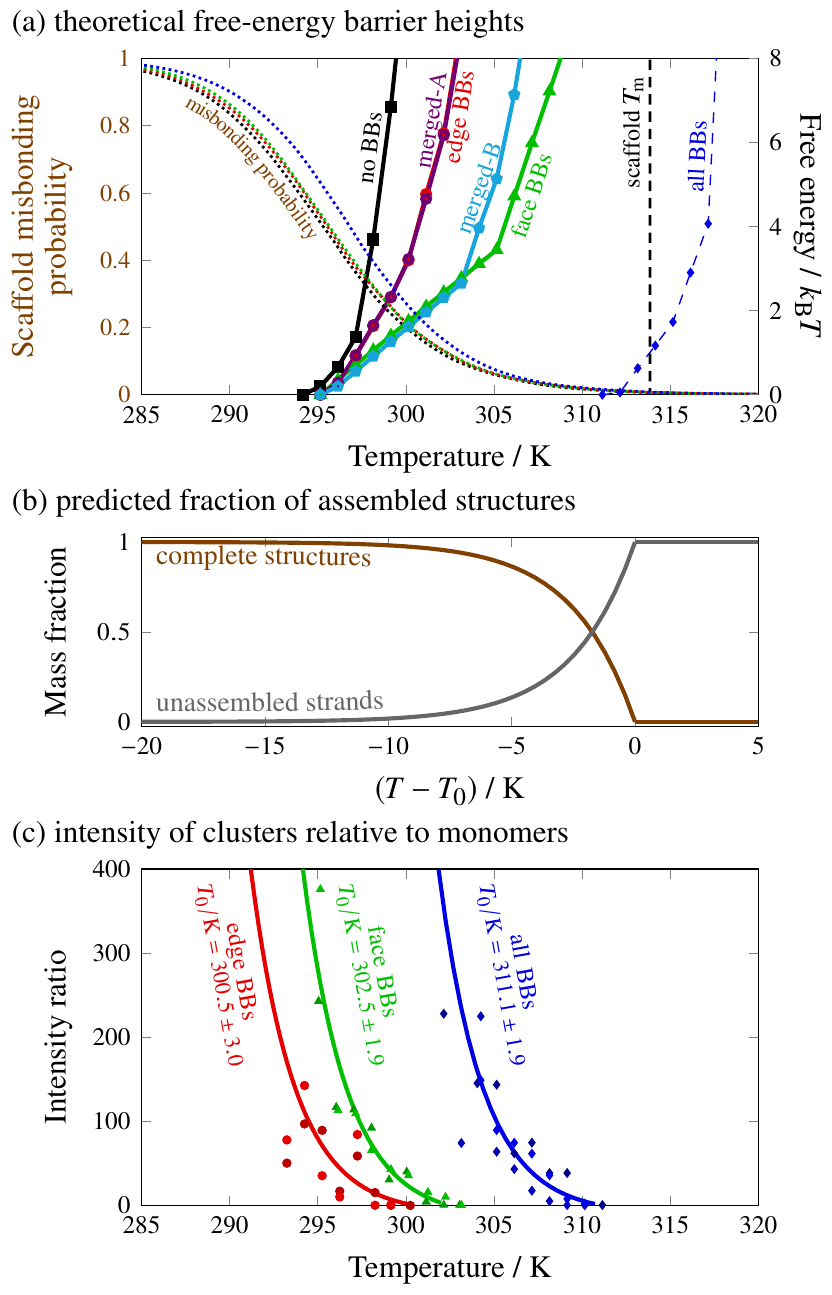}
\vskip1.25ex
\caption{(a) The height of the nucleation barrier, at the initial free-strand concentration, as a function of temperature from theoretical calculations.  The all-BB barrier refers to the nucleation of a network of boundary bricks, as discussed in the text.  We also show the predicted scaffold melting temperature, \TM{}, and the median probability that a scaffold strand forms at least one mis-interaction in the absence of successful nucleation.  (b) The predicted evolution of the unassembled-strand and complete-structure populations over the course of a nucleation-limited annealing protocol, determined from the nucleation-barrier calculations (see \SIseczref{subsec-cluster-population-ratios}).  (c) The ratio of the experimentally determined intensities of the two populations as a function of temperature, alongside an exponential fit $c (\exp[-a(T-T_0)]-1)$ for each system, with $1/a = \SI{2.5}{\kelvin}$ (see \SIseczref{subsec-cluster-population-ratios}); points are shown for multiple annealing runs.}
\label{fig-cluster-populations}
\end{figure}

Based on the evidence of high-temperature hybridization (see \SIseczref{subsec-fluorescence}), we hypothesized that the presence of pre-formed clusters involving BBs might play a key role in determining nucleation behavior.
Similar behavior is exhibited in our simulations, where BB dimers form nearly completely prior to structure nucleation (\SIfigzref{fig-cuboid-lowConc-clustBonds}).
We further tested this idea by running simulations in which BB dimers were merged into permanently bonded units, mimicking the result of high-temperature hybridization in the experimental system.
To this end, simulations with merged edge BBs (`merged-A'; see \SIfigzref{fig-BB-illustrations}) confirmed that nucleation in this system is analogous to the edge-BB structure (\figrefsub{fig-cluster-size-analysis}{c}).

This hypothesis is supported by free-energy calculations using a discrete combinatorial model~\cite{Jacobs2015,Jacobs2015b,Jacobs2015c}, in which each distinct subunit type is represented as a node in an abstract graph that describes the connectivity of the target structure.
Assuming that all 16- and 24-bp domains hybridize completely at high temperatures, we merged the corresponding pairs of subunits to account for changes in the local subunit connectivity due to the incorporation of each type of boundary brick.
We then used this model to calculate the free-energy barrier to nucleation by further assuming that the number of subunits in a partially assembled cluster is a good reaction co-ordinate (see Methods).
These free-energy calculations predict that the heights of the nucleation barriers, and thus the logarithms of the nucleation rates, vary rapidly with temperature (\figrefsub{fig-cluster-populations}{a}).
Furthermore, the relative ordering of the nucleation-barrier curves for the no-BB, edge-BB, and face-BB systems is consistent with the DLS and simulation results, indicating that merging subunits via high-temperature hybridization is sufficient to modify the nucleation behavior.
For comparison, we show the predicted melting temperature \TM{} below which the scaffold-strand core of the cuboid is thermodynamically stable; the model predicts that successful nucleation always requires that the system be supersaturated by lowering the temperature below the scaffold-strand \TM{}.
We also show that the effects of strand misbonding are captured by a simple estimate of the probability of pairwise mis-interactions (see Methods).
As in our simulations, the misbonding probabilities are nearly independent of boundary-brick incorporation.

Interestingly, we found that the assembly of the all-BB structure follows a three-step mechanism that is not well described by a one-dimensional free-energy landscape.
In this system, pairs of pre-formed multimers can hybridize with one another via multiple 8-bp domains.
Consequently, bonding networks that are dominated by BBs begin to form at temperatures where all single 8-bp hybridizations are unstable, leading to extensive boundary-brick bonding and large cluster-size fluctuations in simulations above \SI{310}{\kelvin} (\SIfigzref{fig-cuboid-lowConc-clustBonds}).
Our simulations show that the nucleation of the interior of the structure then occurs in a separate assembly step, at temperatures slightly below the predicted scaffold-strand melting temperature, \TM{}.
Because the theoretical results assume a one-dimensional order parameter, we only show the predicted free-energy barrier that pertains to the formation of an initial network of boundary bricks in the all-BB system in \figrefsub{fig-cluster-populations}{a}.

\subsection*{Nucleation strongly affects self-assembly yield}

Because our free-energy calculations predict that the height of the nucleation barrier also depends strongly on the subunit concentration, the nucleation rate is expected to decrease with monomer depletion~\cite{Jacobs2016}.
The changing concentration of unassembled subunits is therefore predicted to result in the continued production of complete structures at temperatures below $T_0$ in an annealing ramp where nucleation is rate-limiting (\figrefsub{fig-cluster-populations}{b}; see \SIseczref{subsec-cluster-population-ratios}).
This prediction takes into account the temperature scaling derived from the calculated nucleation barriers and the temperature dependence of the hybridization free energies (\SIfigzref{fig-cuboid-hybridisation-energies}), assuming perfect stoichiometry and zero aggregation.
To test this prediction, we integrated the experimentally determined total scattering intensity associated with each peak in the cluster-size distribution and, assuming that this intensity is proportional to the number density, determined the ratio of large- to small-\Rh{} populations.
The trends shown in \figrefsub{fig-cluster-populations}{c} for the edge-BB, face-BB and all-BB structures follow the predictions of our free-energy calculations, as the intensity ratios are consistent with the functional form and temperature scaling shown in \figrefsub{fig-cluster-populations}{b}.
Since there must be some leftover subunits due to imperfect stoichiometry (measured to be approximately \SI{\pm10}{\percent}), we did not expect the unassembled population to decay to zero in the experimental system.
However, the associated intensity did not attain a constant level before the small-\Rh{} peak fell below the detection range of the instrumentation.

These observed variations in nucleation behavior therefore provide a likely explanation for the extreme differences in yields among our structural variants and the similarity between the ranking of the final yields and the order of the initial assembly transitions.
At any given temperature, only a fraction of the potential structures ultimately form because nucleation slows as large clusters are produced; consequently, decreasing the temperature by means of an annealing protocol is necessary to continue driving nucleation of additional structures.
However, our simulations and no-BB DLS measurements indicate that misbonding dominates below \SI{295}{\kelvin}.
Structure designs that nucleate at higher temperatures thus benefit from a broader temperature range over which nucleation can occur.

\subsection*{Nucleation behavior and kinetic stability can be independently tuned}

\begin{figure}[t!]
\includegraphics{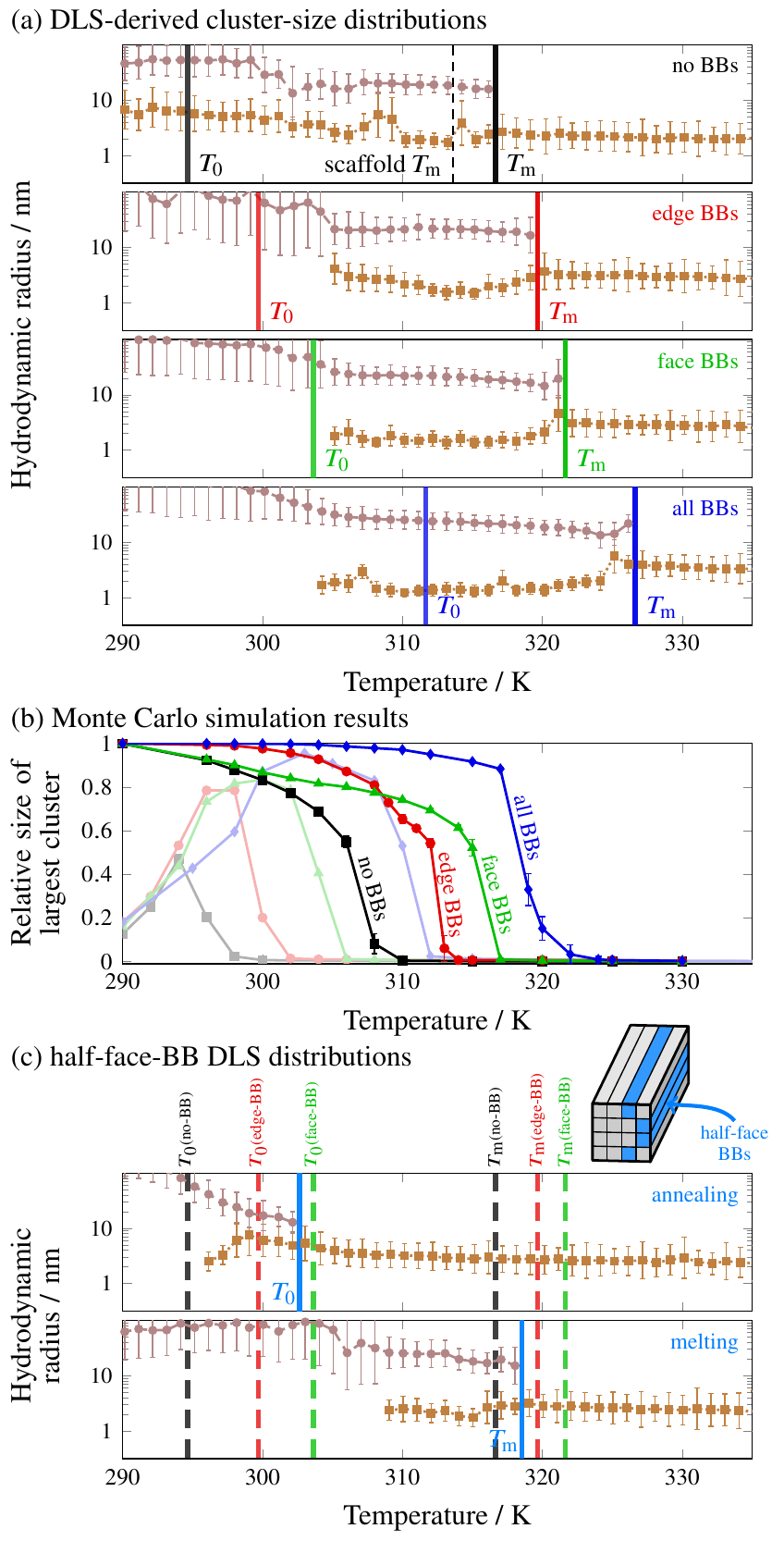}
\caption{(a) Cluster-size distributions determined from DLS melting experiments (cf.~\figrefsub{fig-cluster-size-analysis}{a}).  The temperatures at which all structures have melted, \TM{}, are compared to the initial assembly temperatures, $T_0$.  The no-BB structures melt completely at a temperature close to the predicted scaffold-strand \TM{} (dashed line).  (b) The hysteresis in the size of the structure as a function of temperature can be seen by comparing the long-time-limit average cluster sizes in simulations initialized from pre-assembled structures (solid lines) and from an unassembled solution (faded lines from \figrefsub{fig-cluster-size-analysis}{c}).  In (c), we show the annealing (top) and melting (bottom) cluster-size distributions for a half-face BB system.  The inset shows a schematic illustration of the BB locations.  The temperatures $T_0$ and \TM{} for the no-BB, edge-BB, and face-BB systems are shown by dashed vertical lines for comparison.  Importantly, the order in which the half-face-BB and edge-BB structures nucleate is different from the order in which they melt.}\label{fig-melting}
\end{figure}

The differences among our cuboid variants do not affect the thermodynamic properties of the scaffold strands, which comprise the bulk of the structure.
However, incorporating boundary bricks can, in principle, increase the kinetic stability of assembled structures.
To examine this effect, we reversed the temperature ramp and used DLS to track the melting of assembled structures.
We observed the complete melting of all structures in solution, as evidenced by the disappearance of the $\Rh\sim\SI{20}{\nano\metre}$ population, at considerably higher temperatures than the assembly transitions (\figrefsub{fig-melting}{a}).
The complete melting transitions, \TM{}, of the cuboid variants occurred in the reverse order of the assembly transitions, $T_0$, indicating that a strong bonding network of boundary bricks provides a kinetic barrier to disassembly.
However, the differences in melting temperatures were generally smaller for structures that nucleate at lower temperatures, suggesting that the boundary bricks affect disassembly to a lesser extent than they affect nucleation.

Melting simulations of fully formed structures show similar trends (\figrefsub{fig-melting}{b}).
Analysis of the simulation trajectories reveals that scaffold bricks at the edges of the no-BB and face-BB structures disassemble first.
The face-BB structures therefore lose bricks at lower temperatures than the edge-BB structures, although the face boundary bricks provide a larger barrier to complete disassembly.
Disassembly occurs most abruptly in the case of the all-BB structures, with bricks initially dissociating from the unprotected ends of the structure.
Consistent with the assembly simulations, the all-BB structures disassemble via a three-step disassembly mechanism, in which large networks of boundary-brick dimers persist for a few degrees above the apparent melting temperature (\figrefsub{fig-melting}{b}).

To distinguish between the effects of nucleation and kinetic stability, we designed the `half-face-BB' cuboid shown schematically in \figrefsub{fig-melting}{c}.
By incorporating face BBs on only one half of the structure, we predicted that we would see improved nucleation behavior, as with the full face-BB structure, but reduced kinetic stability.
DLS confirmed that this structure initially nucleates at a temperature close to the face-BB $T_0$ (\figrefsub{fig-melting}{c}), in agreement with our simulations and free-energy calculations (\SIfigzref{fig-face-halfface}).
The assembly yield (\SIfigzref{fig-sample-afm-comb-zoomed-half}) is dramatically improved relative to the no-BB structure, but is less than that of the face-BB structure, presumably because one half of the cuboid is not protected by boundary bricks and is thus more susceptible to aggregation from low-temperature misbonding.

Importantly, DLS reveals that the half-face-BB structure melts \textit{before} the edge-BB structure does, implying that the lack of boundary-brick protection on one face facilitates disassembly of the complete structure.
Comparing the half-face-BB and no-BB systems, which have similar melting temperatures, highlights the crucial role of enhanced nucleation, as opposed to increased stability, in improving the yield.
More generally, this example demonstrates that the nucleation behavior and thermal stability of DNA-brick nanostructures can be independently tuned.

\subsection*{Nucleation pathways are determined by the connectivity of pre-formed clusters}

\begin{figure}[t!]
\includegraphics{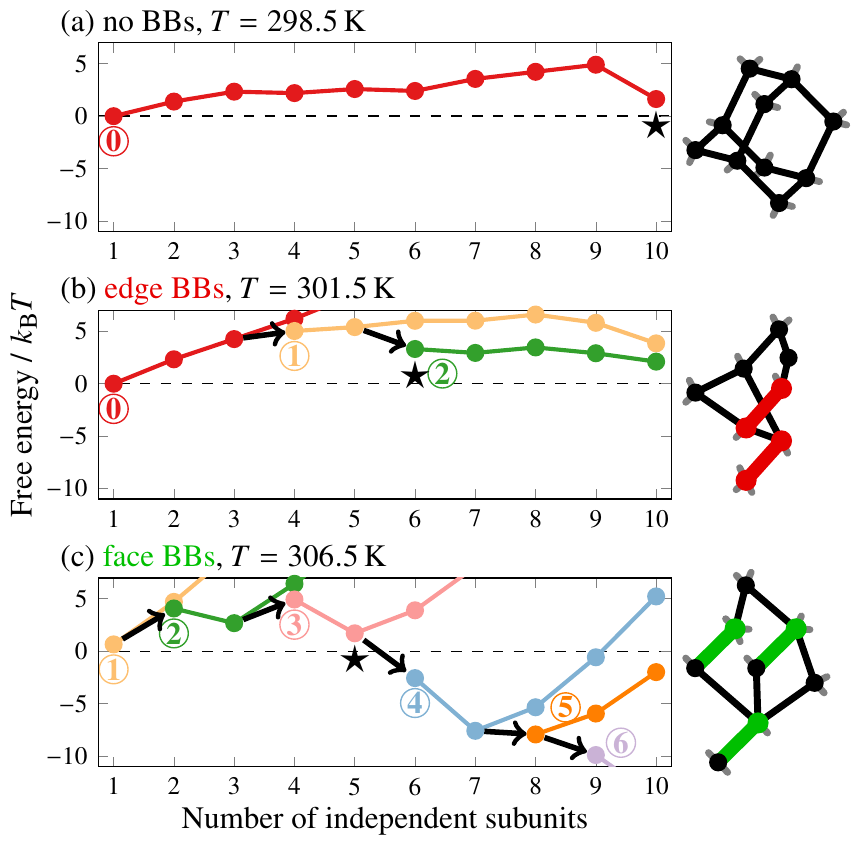}
\caption{The free energy as a function of cluster size and composition from theoretical calculations, showing that the minimum-free-energy self-assembly pathway depends on the presence of pre-formed multimers.
  The number of pre-formed multimers within each cluster is indicated by the ringed numbers and the corresponding colors.
  Clusters on the minimum-free-energy pathway grow by incorporating one independent subunit at a time, which may be either a single scaffold brick (following a colored line) or a pre-formed multimer (following an arrow).
  For ease of comparison, each free-energy landscape is shown at a temperature at which the nucleation barrier is approximately $5\,k_\text{B}T$.
  The post-critical nuclei, which coincide with the first subunit addition after the highest point on the minimum-free-energy path, are indicated by stars.
  Schematic diagrams of representative post-critical nuclei, with pre-formed multimers indicated by colored lines, are shown on the right.
  For a comparison with simulation trajectories, see \SIfigzref{fig-sim-pathways}; representative pathways are also illustrated in Figs~\zref{fig-minfe-pathways-noBB}--\zref{fig-minfe-pathways-faceBB}.}
\label{fig-pathways}
\end{figure}

To identify the microscopic origin of the differences in nucleation behavior, we calculated minimum-free-energy pathways using our theoretical model (\figref{fig-pathways}).
For each structure, we determined the free-energy as a function of the number of independent subunits and the number of pre-formed dimers at a temperature where the nucleation barrier is approximately $5\,k_\text{B}T$, which is comparable to the barrier height at which nucleation was observed in previous simulations of this model~\cite{Reinhardt2014,Jacobs2015b}.
The typical order in which dimers and scaffold strands are incorporated into a growing cluster is indicated by the minimum-free-energy nucleation pathways in \figref{fig-pathways} and illustrated in Figs~\zref{fig-minfe-pathways-noBB}--\zref{fig-minfe-pathways-faceBB}.
Importantly, these calculations allow us to identify typical post-critical nuclei, the smallest multi-strand clusters that are more likely to grow via strand addition than to dissociate and whose formation is thus the rate-limiting step on each predicted nucleation pathway.
The topologies of these clusters are shown in \figref{fig-pathways}; however, because there are many topologically equivalent clusters within each structure, with unique sequences for the hybridized segments, there are numerous post-critical nuclei comprising distinct strands with slightly different free energies.

These landscapes reveal crucial differences between the edge-BB and face-BB structures, which contain the same number of 48-nt BBs, and point to the key role of the connections between the pre-formed dimers and interior scaffold strands.
Topologically, this difference arises from the fact that face-BB dimers contain segments that directly connect to the fully-interior scaffold strands, whereas the edge-BB dimers are only indirectly connected to these `core' strands (see \SIfigzref{fig-BB-illustrations}).
Because each subunit addition results in a loss of translational entropy, the free energy on a nucleation pathway only decreases when multiple 8-bp bonds are formed with a single subunit addition, resulting in a topologically closed cycle.
In the absence of boundary bricks, our previous work has shown that the post-critical nucleus at typical nucleation temperatures is a tricyclic cluster comprising twelve 8-bp bonds and ten subunits~\cite{Jacobs2015}.
Yet by incorporating pre-formed dimers, fewer independent subunits are needed to reach a post-critical nucleus.
The 8-bp bonds can thus be weaker, leading to an elevated nucleation temperature.
Despite the fact that the edge-BB and face-BB dimers have the same number of 8-nt domains for binding to other subunits, the topologies of the minimum-free-energy clusters in these structures are different: the edge-BB structures require six subunits, including two BB dimers, to form a post-critical bicyclic cluster (\figrefsub{fig-pathways}{b}), while face-BB structures only require five subunits, including three dimers (\figrefsub{fig-pathways}{c}).
Consistent with the predicted pathways, simulation trajectories show that BB dimers comprise a larger fraction of the post-critical clusters in the face-BB structure than in the edge-BB structure (\SIfigzref{fig-sim-pathways}).
We can also exclude the concentration of pre-formed dimers as the determining factor by comparing the half-face-BB and edge-BB structures, as these systems contain the same number of pre-formed dimers but nucleate at significantly different temperatures.

Based on these findings, we hypothesized that by changing the local connectivity of the edge BBs, we might be able to reproduce the enhanced nucleation behavior of the face-BB structure.
To test this hypothesis directly, we ran simulations of the edge-BB system in which we explicitly merged each edge dimer with one of its neighboring scaffold strands in the target structure (\SIfigzref{fig-BB-illustrations}).
The only difference between this structure (`merged-B'; see \SIfigzref{fig-BB-illustrations}) and a normal edge-BB dimer is that this additional connection to the interior scaffold strands, which would otherwise need to form spontaneously during the nucleation process and thus entail a loss of translational entropy, has been fixed in place.
This modification leads to post-critical nuclei that comprise five independent subunits, resulting in assembly behavior that is nearly analogous to that of the face-BB structure (Figs~\refSub{c}{fig-cluster-size-analysis} and \refSub{a}{fig-cluster-populations}).
Thus, although this particular modification would be difficult to achieve experimentally using DNA bricks, our simulations and theory show that the addition of a single connection to the interior of the structure can alter the nucleation behavior significantly.

\subsection*{Design rules for enhanced nucleation}

Based on our experimental, simulation and theoretical findings, we propose four design rules for enhancing the nucleation behavior and assembly yield in addressable systems:
\begin{enumerate}[leftmargin=1em]
\setlength\itemsep{-0.25\baselineskip}
\item The key determinant of the structure yield is the separation between the initial nucleation and misbonding temperatures.
  While the misbonding temperature is set by the pairwise interactions and the subunit concentrations, the nucleation temperature can be tuned through rational structure design.
  By contrast, changes to the subunit interactions that uniformly affect both correct and incorrect bonds are unlikely to improve the yield.
\item Altering the `valency' of specific subunits to create multi-step pathways, for example by forming boundary-brick dimers at high temperatures, is a viable strategy for controlling nucleation, because it can change the number of independent subunits in the critical nucleus.
  On the other hand, tuning individual bond strengths is a less effective strategy for selecting a specific nucleation pathway, since the number of parallel pathways grows superextensively with the size of the target structure.
\item Controlling the topology of the critical nuclei is crucial.
  It may not be optimal simply to add high-valency subunits, as in the case of the edge BBs.
  Instead, efficient nucleation requires that the critical nuclei contain many stabilizing bonds but few subunits, favoring the formation of free-energy-reducing topological cycles earlier in the nucleation pathway.
  This is achieved in the case of the face-BB and merged-B structures by maximizing the number of bonds between the pre-formed dimers and the interior scaffold bricks.
\item Only a small portion of a structure needs to be optimized to achieve enhanced nucleation behavior.
  For example, comparison of the half-face-BB and no-BB systems shows that modifying fewer than \SI{20}{\percent} of the subunits drastically raises the initial nucleation temperature and markedly improves the yield.
\end{enumerate}

\section*{Discussion}

By combining dynamic light scattering with a coarse-grained theoretical model, we have shown that the ultimate yield of correctly assembled structures is largely determined by the nucleation pathway.
As a specific example, we have investigated the role of nucleation kinetics in addressable self-assembly by modifying the bonding characteristics of specific subunits at the boundaries of a DNA-brick nanostructure.
We have shown that the location and design of the altered subunits determine the free-energy landscape for self-assembly and control the temperature at which nucleation first becomes feasible.
Moreover, the nearly quantitative agreement between the predictions of a coarse-grained model and our experimental results allows us to rationalize these striking effects on the self-assembly behavior.

Taken together, our experiments and modeling establish practical design principles for improving the self-assembly of addressable nanostructures.
In a typical annealing protocol, structures have a limited time and temperature window in which to form: at high temperatures, a large free-energy barrier inhibits nucleation, while at low temperatures, self-assembly is limited by kinetic arrest.
The key to successful self-assembly is to increase the width of the temperature window over which nucleation can occur, thereby maximizing the thermodynamic segregation between the critical nucleation step and detrimental misbonding.
This can be achieved by stabilizing the critical nuclei, which allows self-assembly to proceed when the subunit interactions are still relatively weak.
To demonstrate this principle with DNA bricks, we have shown that the increased valency of boundary-brick dimers, which assemble at temperatures much higher than those at which nucleation can occur, lowers the free-energy barrier to nucleation by decreasing the entropic cost of forming a critical number of stabilizing bonds.
However, this strategy only works if the high-valency subunits are optimally connected to the remainder of the structure, as evidenced by the difference in nucleation behavior between the edge-BB and face-BB structures.
More generally, our results show that it is possible to use a relatively small number of high-valency subunits to design the nucleation pathway rationally and suggest that this approach is not necessarily limited to manipulating bricks at the boundaries of a structure.

Our experiments provide the first explicit characterization of three-dimensional structure nucleation in the context of addressable self-assembly.
This advance has been enabled by our use of DLS, which allows us to probe multi-strand structure growth, as opposed to the fraction of inter-subunit bonds that are formed at a given temperature.
This distinction is particularly evident in the system evaluated here, where the initial nucleation temperature does not necessarily correlate with the maximal increase in DNA base-pairing.
Furthermore, the cluster-size distributions that we obtain from DLS resolve the populations of unincorporated bricks, complete structures and aggregates, making it possible to track the evolution of these species throughout the course of an annealing protocol.
Together with the complementary AFM-based validation, these measurements provide experimental evidence that DNA bricks self-assemble via a nucleation-and-growth mechanism and reveal the relationship between the design of addressable structures and their nucleation kinetics.

The excellent agreement between the predictions of our theoretical model and our experimental results demonstrates that our coarse-grained approach captures the fundamental physics of addressable self-assembly.
This agreement gives us confidence that our theory and simulations can be used to guide rational design strategies for complex self-assembly, not only in the context of DNA bricks specifically, but --- precisely because of the generality of the models used --- also for optimizing addressable systems more broadly.
We anticipate that the principles established here will therefore guide efforts to design the nucleation behavior of colloidal systems such as supramolecular and nanoparticle lattices~\cite{Macfarlane2011,Huang2015,Lin2017}, protein nanostructures~\cite{Bale2016} and DNA-origami-based systems with programmable interactions~\cite{Gerling2015}.
For example, analogous pre-nucleation clusters could be constructed by forming high-temperature bonds between caged nanoparticles~\cite{Liu2016} or by directly introducing a small population of dumbbell-like subunits.
Alternatively, the connectivity of specific subunits could be altered by changing the arrangement of directional patches on colloidal particles~\cite{Wang2012}.
As we have demonstrated here, successful implementation will require knowledge of the effects of such modifications on the critical nucleus for structure assembly, which dictates the optimal design strategy for any specific system.

\section*{Materials and Methods}

In the Supplementary Material, we describe how we chose the DNA sequences for the strands for each system studied. We also provide complete details of the annealing protocols, the conditions used in AFM and gel electrophoresis, and the protocols used when obtaining fluorescence and light scattering data in the Extended Methods. Supporting data are available at the \href{https://doi.org/10.17863/cam.22991}{University of Cambridge Data Repository}~\cite{supporting-data}.

\subsection*{Structure annealing}
Structures were assembled using a strand concentration of \SI{153}{\nano\Molar} per sequence in a buffer of \SI{15}{\milli\Molar} \ce{MgCl2}, \SI{0.5}{\milli\Molar} EDTA and \SI{5}{\milli\Molar} Tris at pH 8.
Strands in the reaction mixture were denatured at \SI{90}{\celsius} for \SI{10}{\minute} and then gradually cooled via either (i) a \SI{15.2}{\hour} protocol (reciprocal cooling rate \SI{12}{\minute\per\kelvin}) or (ii) a \SI{66}{\hour} protocol (reciprocal cooling rate \SI{52}{\minute\per\kelvin}).

\subsection*{Atomic force microscopy}
Samples from annealing protocol (ii) were immobilized for \SI{10}{\minute} on poly-\textsc{l}-ornithine coated mica discs and imaged in liquid in intermittent contact mode using a BioLever Mini cantilever and JPK Nanowizard 3 AFM.

\subsection*{Agarose gel electrophoresis}
Structures were analyzed via gel electrophoresis on a gel made from \SI{2}{wt\%} agarose in $0.5\times \text{TBE}$ and \SI{10}{\milli\Molar} \ce{MgCl2}. Electrophoresis was performed at \SI{80}{\volt} and \SI{4}{\celsius} for \SI{2}{\hour}. The gel was post-stained with ethidium bromide and the yield was estimated using GelBandFitter software~\cite{Mitov2009}.

\subsection*{Fluorescence annealing}
Annealing protocol (i) was used for fluorescence annealing experiments with \SI{10}{\nano\Molar} SYBR green~I solution~\cite{Zipper2004} added to the reaction mixture. The fluorescence signal was measured as a function of temperature with an ABI Prism 7900HT-Fast Real Time PCR system at \SI{488}{\nano\metre}.

\subsection*{Static and dynamic light scattering}
Using annealing protocol (i), light scattering measurements were performed in the last \SI{2}{\minute} of each temperature step.
Light scattering of \SI{20}{\micro\liter} samples was measured using a Malvern Zetasizer NanoZSP apparatus at an angle of \ang{173}.
For DLS, the intensity auto-correlation function was computed from 12 measurements at \SI{10}{\second} intervals.
Cluster-size distributions were determined from the auto-correlation data using multiple regularization methods~\cite{Hansen2017} to verify their robustness (\SIfigzref{fig-DLS-regularisation}).

\subsection*{Reference hydrodynamic radius calculations}
The hydrodynamic radius of a freely jointed chain is $R_\text{h}=(3\uppi N \ell^2/128 )^{1/2}$~\cite{Teraoka2002}, where $N$ is the number of segments and $\ell$ is the length of a segment. 
To estimate the hydrodynamic radius for single-stranded DNA, we used a typical Kuhn length of $\ell=\SI{4.45}{\nano\metre}$ and length per DNA base of $b=\SI{0.676}{\nano\metre}$~\cite{Chi2013}.
A 32-nt scaffold brick comprises $N \approx 32\times b/\ell=4.9$ Kuhn segments and hence $R_\text{h} \simeq \SI{2.7}{\nano\metre}$; for a 48-nt boundary brick, $R_\text{h} \simeq \SI{3.3}{\nano\metre}$. 
Since the quantities $\ell$ and $b$ used here do not correspond to the temperatures and salt concentrations used in our experiments, these calculations only provide us with rough estimates of the magnitudes of \Rh{} for unhybridized strands.

For cylindrical structures, the translational diffusion coefficient is~\cite{LaTorre1981}
\begin{equation}
D_\text{tr} = \frac{k_\text{B}T \left[\ln(L/d)+\gamma(d/L)\right]}{3\uppi\eta L} ,\label{eq-latorre}
\end{equation}
where $k_\text{B}$ is the Boltzmann constant, $T$ is the absolute temperature, $\eta$ is the viscosity of the medium, $L$ is the cylinder length, $d$ is the cylinder diameter, and $\gamma$ is an end-effect correction given by $\gamma(x) = 0.312+0.565x+0.1x^2$.
Assuming that the hydrodynamics of a cylinder are well approximated by a sphere with hydrodynamic radius $R_\text{h}$, we can equate this diffusion coefficient to that of a sphere using the Stokes--Einstein--Smoluchowski equation,
\begin{equation}
D_\text{tr} = \frac{k_\text{B}T}{6\uppi\eta R_\text{h}},\label{eq-stokes}
\end{equation}
leading to an approximate hydrodynamic radius of
\begin{equation}
  R_\text{h} = (L/2)\left[\ln(L/d)+\gamma(d/L)\right]^{-1}.
\end{equation}
Assuming a typical interhelical spacing of $\sim\!\SI{2.5}{\nano\metre}$~\cite{Fischer2016}, our target structure can be treated as a cylinder with circumscribed diameter $d\approx \SI{15}{\nano\metre}$ and length $L\approx \SI{86}{\nano\metre}$, resulting in an expected hydrodynamic radius of $R_\text{h} \simeq \SI{20}{\nano\metre}$.

\subsection*{Image analysis}
Particle identification in AFM images was done by applying the threshold function of Gwyddion version 2.5.0~\cite{Necas2012}.
The reference \Rh{} distribution shown in \figrefsub{fig-DLS-validation}{a,ii} was calculated using the lengths, $L$, and aspect ratios, ${L/d \simeq L^2/A}$, where $A$ is the projected area, of the particles assuming the cylindrical formula given above; particles with a minimum width greater than \SI{25}{\nano\metre}, which correspond to overlapping structures, were excluded from this calculation.
To compute the distribution function, particles were weighted by $\Rh^6$, and the distribution was normalized.

To calculate the fraction of correctly assembled structures in an AFM image, we selected all particles that satisfied constraints on both the projected area, ${\SI{450}{\nano\metre^2} \le A \le \SI{1500}{\nano\metre^2}}$, and the circularity, ${0.145 \le 4A/\uppi L^2 \le 0.375}$.
These limits were chosen based on the distribution of imaged particles in the purified all-BB system (\SIfigzref{fig-AFM-all-BB-positive-control}).
All particles were weighted by their volumes, as determined by the Laplacian background basis feature of Gwyddion, in order to assess the total fraction of the material contained in the selected particles.
To reduce background noise, we also required that the area of the selected particles measured at half the particle height be at least $0.3A$ and the average height be at least \SI{2}{\nano\metre}.
Standard errors were assigned to the yields by assuming a Poisson distribution based on the calculated yield and the absolute number of selected particles.

\subsection*{Monte Carlo simulations}
We performed lattice Metropolis Monte Carlo simulations of DNA brick self-assembly using a coarse-grained potential and dynamics that preserve the cluster-size dependence of the diffusion rates~\cite{Reinhardt2014, Reinhardt2016, WaymentSteele2017}.
Every DNA brick was represented as a `patchy particle' with four patches corresponding to its four domains, each of which was assigned a specific unique sequence, chosen randomly but with the constraint that patches that are bonded in the target structure have complementary DNA sequences. 
The interaction energies correspond to the hybridization free energies of these sequences obtained from the SantaLucia parameterization~\cite{SantaLucia2004}.
When computing these hybridization free energies, we used a salt correction~\cite{Koehler2005} corresponding to salt concentrations of $\ce{[Na^{+}]}=0$ and $\ce{[Mg^{2+}]}=\SI{0.015}{\Molar}$.  
In the simulations reported here, we used a system of 550 bricks in a box with lattice parameter $150a$, where $a\sqrt{3}$ is the shortest possible distance between any two particles. 
Assuming typical brick dimensions of $(a\sqrt{3})^3\approx\SI{2.5x2.5x2.7}{\nano\metre}$~\cite{Ke2012, Reinhardt2016b}, this set-up corresponds to a concentration of \SI{153}{\nano\Molar}. 
We accounted for boundary bricks by imposing rigid bonds between dimers (or, in certain cases, larger multimers) of these patchy particles that would be merged into a single boundary brick in experiment~\cite{WaymentSteele2017}.
Particles connected in this way remain at a fixed distance and dihedral angle to one another throughout the simulation. Non-interacting patches on the outside of the target structure were assigned poly-T sequences.
We estimated a Kirkwood-like~\cite{Kirkwood1954} hydrodynamic radius of each cluster by computing
\begin{equation}
\Rh{}\approx \frac{N(N-1)}{2} \left[\sum_{\langle ij\rangle} \frac{1}{r_{ij}} \right]^{-1} + 0.5\times \SI{2.7}{\nano\metre},
\label{eq-sim-Rh}
\end{equation}
where $N$ is the number of particles in the cluster, $\langle ij\rangle$ indicates a summation over every pair of particles $i$ and $j$ in the cluster, and $r_{ij}$ is the distance between them on the lattice, using the typical brick dimensions given above to determine that $a\approx \SI{1.48}{\nano\metre}$ for the lattice unit of length.
The monomer \Rh{} is set to \SI{2.7}{\nano\metre}, while the addition of $0.5\times \SI{2.7}{\nano\metre}$ in \eqref{eq-sim-Rh} crudely accounts for the dangling ends of monomers at one of the bases of the structure, which are not otherwise accounted for in the coarse-grained model.
\eqref{eq-sim-Rh} thus predicts that an scaffold brick dimer will have a hydrodynamic radius of ${\Rh \simeq \SI{3.9}{\nano\metre}}$.

\subsection*{Free-energy calculations}

All free-energy calculations were carried out using the abstract-graph model described in Ref.~\citenum{Jacobs2015}.
The free energy of a particular cluster $g$, comprising a set of subunits $\mathcal{V}(g)$, is
\begin{equation}
  \frac{F_g}{k_{\text{B}}T} = -\!\!\!\sum_{\substack{i,j\in\mathcal{V}(g)\\j\in\mathcal{E}(i)}}\!\! \frac{\varepsilon_{ij}}{2} - (N_g - 1) \ln\frac{\rho}{q_{\text{rot}}} + (N_g - B_g - 1) \ln q_{\text{dih}},
\end{equation}
\vskip-0.5ex
\noindent where $N_g$ is the number of subunits in the cluster, ${\mathcal{E}(i)}$ indicates the set of strands that are neighbors of strand $i$ in the target structure, and the dimensionless bond strengths are ${\varepsilon_{ij} = \ln \mleft[\exp\mleft(-\upDelta G_{ij}/k_{\text{B}}T\mright) - 1\mright]}$.
We determined $\upDelta G_{ij}$ for each pair of complementary sequences $i,j$ in the experimental systems using the SantaLucia parameterization described above.
Each subunit, with concentration $\rho$, was assumed to have ${q_{\text{rot}} = 4}$ rotational degrees of freedom, and each single bond was assumed to have ${q_{\text{dih}} = 3}$ dihedral degrees of freedom; these values were chosen to match the Monte Carlo simulations.
$B_g$ refers to the number of `bridges' in the graph~$g$~\cite{Jacobs2015}.
The cluster free energy as a function of the number of correctly bonded subunits is
\begin{equation}
  F(N) = -k_{\text{B}}T \ln \sum_g \boldsymbol{1}\mleft(N_g{=}N\mright) \exp(-F_g/k_{\text{B}}T),
\end{equation}
\vskip-0.75ex
\noindent where ${\boldsymbol{1}(\cdot)}$ is the indicator function.
${F(N)}$ was calculated using the efficient Monte Carlo approach described in Ref.~\citenum{Jacobs2015}.
Similarly, in \figrefsub{fig-pathways}{a}, the cluster free energy was calculated as a function of the total number of subunits and the number of pre-formed dimers.

The melting temperature $T_{\text{m}}$ of an infinite lattice of scaffold strands with co-ordination number ${z = 4}$ was estimated based on the mean of the 8-bp scaffold-strand hybridization free energies by solving the equation
\begin{equation}
  (z/2) \overline{\upDelta G(T_{\text{m}})} = k_{\text{B}} T_{\text{m}}\ln \mleft(\rho / q_{\text{rot}} q_{\text{dih}}\mright).
\end{equation}

\subsection*{Misbonding calculations}

We used a two-state model (i.e.~bonded or not bonded) to calculate the probability that a strand forms at least one misinteraction, assuming that no domains are correctly hybridized.
We found the longest complementary subsequence for each pair of strands $i$ and $j$ that are not neighbors in the target structure and calculated the associated hybridization free energy, ${\upDelta G_{\text{min},ij}}$.
[In cases where there are multiple complementary subsequences $\{s\}$ of the same length for a given pair of strands, we calculated the Boltzmann-weighted sum, ${\upDelta G_{\text{min},ij} = -k_\text{B}T \ln \sum_s \exp(-\upDelta G_s/k_\text{B}T)}$.]
We then computed the probability that a strand $i$ forms a misinteraction, $p_{\text{mis},i}$,
\vskip-2.5ex
\begin{align}
  p_{\text{mis},i} &= Z_{\text{mis},i} / \left(1 + Z_{\text{mis},i}\right), \\
  Z_{\text{mis},i} &= \!\sum_{j \notin \mathcal{E}(i)}\! \rho \exp\mleft(-\upDelta G_{\text{min},ij}/k_\text{B}T\mright).
\end{align}
When computing the probability of scaffold-strand misbonding, the index $i$ represents a scaffold strand, while the index $j$ runs over all strands in the system.
This approximate approach captures the competition between designed and incorrect bonding seen in the simulations (\figrefsub{fig-cluster-size-analysis}{c}) remarkably well.
\vskip-1.5ex

\begin{acknowledgments}
We thank Daan Frenkel for helpful discussions. This work was supported by the Engineering and Physical Sciences Research Council [Programme Grant EP/I001352/1], the European Regional Development Fund [100185665], Fraunhofer Attract Funding [601683] and the National Institutes of Health [Grant F32GM116231].
\end{acknowledgments}

\renewcommand{\doibase}{https://dx.doi.org/}

\clearpage

\makeatletter
\renewcommand*{\thesection}{\arabic{section}}
\renewcommand*{\thesubsection}{\thesection.\arabic{subsection}}
\renewcommand*{\p@subsection}{}
\renewcommand*{\thesubsubsection}{\thesubsection.\arabic{subsubsection}}
\renewcommand*{\p@subsubsection}{}
\renewcommand\@seccntformat[1]{\csname the#1\endcsname\quad}

\renewcommand{\secref}[1]{Sec.~SI-\ref{#1}}

\makeatother
\onecolumngrid
\renewcommand{\thepage}{S\arabic{page}}
\setcounter{page}{1}

\section*{Supplementary Information}
\vskip0.5ex
\twocolumngrid

\section{Extended Methods}
\renewcommand*{\theHfigure}{\thepart.\thefigure}
\renewcommand{\thefigure}{S\arabic{figure}}
\setcounter{figure}{0}

If not otherwise noted, chemicals were purchased from Sigma Aldrich Chemie GmbH (Munich, Germany).

\subsection{Annealing protocols}\label{subsec-annealing}
DNA sequences for the target cuboid structure are given in Sec.~SI-\ref{sec-seqs}. For the all-BB system, sequences were taken from Ref.~\citenum{Ke2012} (Table S8), and DNA strands were appropriately decoupled to split the relevant boundary bricks for the face-BB, edge-BB and no-BB systems. All sequences were purchased from Eurofins Genomics in \SI{100}{\micro\Molar} stocks in dd\ce{H2O}, and then pooled using a Tecan Genesis Workstation 150 liquid handling robot. We used a strand concentration of \SI{153}{\nano\Molar} in $1\times\text{assembly}$ buffer, i.e.,~a solution of \SI{15}{\milli\Molar} \ce{MgCl2}, \SI{0.5}{\milli\Molar} EDTA and \SI{5}{\milli\Molar} Tris, pH 8. The strand solution was denatured at \SI{90}{\celsius} for \SI{10}{\minute} and then gradually cooled. We used two linear cooling protocols: (i) in the \SI{15.2}{\hour} protocol, the reciprocal cooling rate was \SI{12}{\minute\per\kelvin}, and (ii) in the \SI{66}{\hour} protocol, it was \SI{52}{\minute\per\kelvin}. The annealed samples were stored at \SI{4}{\celsius}.

Prior to the reference DLS measurement (positive control), the all-BB sample assembled in the \SI{66}{\hour} protocol was supplemented with \SI{2.5}{\milli\Molar} EDTA to reduce high-molecular-weight contamination.
In the context of DLS experiments, we therefore refer to this sample as `purified'.

Prior to the reference AFM imaging (positive control), the all-BB sample assembled in the \SI{66}{\hour} protocol was ultrafiltered using Amicon (UFC510024, Millipore, Merck KGaA, Darmstadt, Germany) filter units, with a molecular weight cut-off of \SI{100}{\kilo\dalton}, to reduce the fraction of small particle contamination and to improve the image quality.  To this end, the assembled sample was mixed with pre-chilled $1\times\text{assembly}$ buffer to the maximum admitted volume and centrifuged for \SI{10}{\minute} at $\num{14000}g$ at \SI{4}{\celsius}.  Subsequently, the flowthrough was discarded, and the filter unit was loaded with buffer and centrifuged again.  This process was repeated three times in total.  Finally, the concentrated filtrate was eluted at $\num{1000}g$ for \SI{2}{\minute} at \SI{4}{\celsius}.

\subsection{Atomic force microscopy}\label{subsec-afm-protocols}
Samples were prepared following the \SI{66}{\hour} annealing protocol. A freshly cleaved mica disc was coated with \SI{100}{\micro\liter} of \SI{0.5}{wt\%} poly-\textsc{l}-ornithine solution for \SI{5}{\minute} and rinsed three times with  $1\times\text{assembly}$ buffer. In order to be able to image samples in liquid mode, an acrylic glass ring was glued by Thin Pour (Reprorubber) onto a slide to surround the mica disc and form a fluid cell. For each sample, \SI{1.5}{\pico\mole} per brick was deposited on the coated mica for \SI{10}{\minute}. Afterwards, the cell was filled with $1\times\text{assembly}$ buffer and imaged using the JPK Nanowizard 3 atomic force microscope and a BioLever Mini cantilever in intermittent contact mode in liquid. Images were recorded with a target amplitude of \SI{15}{\nano\metre}.

Quenching experiments, designed to stop the hybridization reaction at a given temperature during the annealing protocol, were done by immobilizing \SI{5}{\micro\liter} of the reaction mixture on poly-\textsc{l}-ornithine coated and pre-equilibrated mica discs.
As a negative control, the same procedure was performed using a random selection of ssDNA strands that did not contain complementary sequences~\cite{Sajfutdinow2017}.
For both the all-BB structure and the negative control, samples were quenched from \SIlist{318; 314; 311; 308; 300}{\kelvin} during annealing protocol (i) and imaged by AFM, as described above, at ambient temperature.

\subsection{Agarose gel electrophoresis}\label{subsec-gel-electrophoresis}
Assembly of DNA brick structures was confirmed by non-denaturing agarose gel electrophoresis. Samples (\SI{300}{\femto\mole} per brick) were analyzed on a gel made from \SI{2}{wt\%} agarose in $0.5\times \text{TBE}$ and \SI{10}{\milli\Molar} \ce{MgCl2}.  Electrophoresis was performed at \SI{80}{\volt} and \SI{4}{\celsius} for \SI{2}{\hour}. The gel was post-stained with \SI{0.5}{\micro\gram\per\milli\liter} ethidium bromide solution and scanned in using the Intas GDS gel set instrument for structure visualization. To estimate structure yield, the band intensity was approximated by fitting densitometry profiles with an SQP algorithm to Gaussian functions using the GelBandFitter software~\cite{Mitov2009}. The mass of the structure fractions was estimated via a \SI{1}{kb} standard (GeneRuler, ThermoFischer Scientific) and related to the total mass loaded of \SI{850}{\nano\gram}.

\subsection{Static and dynamic light scattering}\label{subsec-scattering}
The same conditions as in the \SI{15.2}{\hour} annealing protocol were used and the measurement was performed in the last \SI{2}{\minute} of the \SI{12}{\minute} cooling step.
\SI{20}{\micro\liter} samples were filled into ZEN2112 quartz cuvettes (Malvern), covered by molecular biology grade mineral oil, and sealed with a plastic lid that was further fixed with tape. 
Light scattering was measured using a Malvern Zetasizer NanoZSP apparatus at an angle of \ang{173}. The viscosity of the samples was determined at five temperatures spanning the region of interest and fitted to $\eta / (\SI{e-5}{\pascal\second}) = 1.78\times \exp[617/(T/\si{\kelvin}-138.5)]$.  
The refractive index was measured to be 1.331. 

For dynamic light scattering, the intensity auto-correlation function was computed from 12 measurements at \SI{10}{\second} intervals.
We interpreted the DLS data in the dilute limit by assuming that all particles diffused independently of one another, since the total strand concentration (approximately \SI{40}{\micro\Molar}) implies that single strands ($\Rh\sim\SI{2.7}{\nano\metre}$) in solution occupied a volume fraction of approximately \SI{0.2}{\percent}.
We further assumed that, after the initial equilibration period of \SI{10}{\minute}, the distribution of cluster sizes remained nearly constant over the DLS measurement period at the end of each temperature step; this assumption is consistent with our observations of rate-limiting structure nucleation.

When analyzing DLS data for solutions comprising a range of particle sizes, the inverse Laplace transform used to obtain a particle size distribution from the intensity auto-correlation function is not uniquely determined~\cite{Provencher1982}, and the choice of fitting functions and parameters can affect the final result~\cite{Hansen2017}.
We have therefore computed multiple fits to the distribution of hydrodynamic radii using several regularization methods, including a range of different smoothing exponents and a maximum entropy constraint (\figref{fig-DLS-regularisation}).
We also performed the analysis using the CONTIN algorithm~\cite{Provencher1982} and via entropy maximization with a uniform prior distribution (not shown).
Although the agreement is not perfect for the individual data points, the trends for the distributions and the fits to a linear combination of Gaussian functions are nearly the same regardless of the regularization procedure used, indicating that the conclusions drawn from the DLS data are robust with respect to the choice of regularization method.
We chose to use the smoothness constraint regularization method recommended in Ref.~\citenum{Hansen2017} with a smoothing exponent of $m=8/5$ (corresponding to \figrefsub{fig-DLS-regularisation}{b}) for all data reported in the main text.

\subsection{Fluorescence annealing}\label{subsec-fluorescence-methods}
In the fluorescence annealing experiments, the same conditions as in the \SI{15.2}{\hour} annealing protocol were used, except that \SI{10}{\nano\Molar} SYBR green~I solution~\cite{Zipper2004} was added to the strand mixture. SYBR green~I in buffer solution was analyzed as a negative control. Samples were placed on a MicroAmp Fast Plate 96-well tray and sealed with adhesive film. The plate was loaded onto the ABI Prism 7900HT-Fast Real Time PCR system, with dye excitation effected by an argon ion laser at \SI{488}{\nano\metre}. The fluorescence signal was detected at \SI{525}{\nano\metre} every \SI{8.5}{\second} and averaged over time at each temperature, and its derivative with respect to temperature was computed numerically. The data were smoothed via a Gaussian filter with a standard deviation of \SI{1.5}{\kelvin}.

\section{Monitoring strand hybridization}\label{subsec-fluorescence}

\subsection{Fluorescence measurements}\label{subsec-fluorescence-measurements}

We monitored the progress of domain hybridization during the annealing protocol via fluorescence, using SYBR green~I as a double-stranded DNA probe (see \secref{subsec-fluorescence-methods})~\cite{Sobczak2012}.
We observed a dominant maximum in the fluorescence derivative between \SI{335}{\kelvin} and \SI{350}{\kelvin} for all structures with boundary bricks (\figrefsub{fig-hybrid}{a}), indicating a significant amount of base-pairing at relatively high temperatures.
However, as we discuss in the main text, no complete structures were assembled at these temperatures.

Comparison with theoretical annealing curves suggests that the assembly of boundary-brick structures is a two-step process.
To demonstrate this, we show in \figrefsub{fig-hybrid}{b} the temperature derivative of the equilibrium number of base pairs in a solution of monomers and dimers~\cite{SantaLucia2004,Koehler2005}, assuming that stable misbonding between non-complementary domains cannot occur (see below).
The high-temperature transitions correspond to the hybridization between pairs of boundary bricks (where continuous 24-bp segments are hybridized) or between one scaffold strand and one boundary brick (with 16-bp hybridized segments).
Consequently, the assembly of the full structure must occur in the presence of these pre-formed clusters.
These calculations also indicate that the fluorescence-signal contributions from each domain length overlap significantly, since the domain melting temperatures vary widely according to their specific sequences, and each hybridization reaction tends to occur over a broad ($\gtrsim$\SI{10}{\kelvin}) range of temperatures.
In particular, the theoretical annealing curves predict a broad maximum associated with the 8-bp domains near \SI{295}{\kelvin}.

Analysis of fluorescence data has previously been used to distinguish between single- and multi-step assembly mechanisms for DNA tile systems with varying domain lengths.
For example, a similar step-wise assembly process was seen in DX-tile structures comprising short (10- and 11-bp) and long (21-bp) hybridizations, and the presence of two distinct maxima in the fluorescence derivative was interpreted as evidence of hierarchical assembly~\cite{Wang2016}.
By contrast, fluorescence measurements of DNA-brick crystallization using equal-length domains exhibited no evidence of hierarchical self-assembly~\cite{Ke2014}.
In our measurements, there appear to be multiple local maxima in the annealing curves at temperatures below the scaffold-strand $\TM \simeq\SI{315}{\kelvin}$, the highest temperature at which our theoretical calculations predict that a lattice of scaffold strands can be thermodynamically stable.
However, these signals are significantly weaker than the higher-temperature hybridizations which dominate the fluorescence signal.
Interpreting the lower-temperature maxima is additionally hindered by several known sources of bias, including high background signals~\cite{Zipper2004} and the preferential binding of SYBR green~I to GC-rich sequences~\cite{Giglio2003}.
Furthermore, the intercalating SYBR green~I probes distort the double-helical structure of DNA molecules~\cite{Suh1995}, which increases their melting temperatures~\cite{Ririe1997} and precludes a quantitative analysis.

\subsection{Hybridization calculations}\label{subsec-hybridization}

All hybridization calculations were carried out using the SantaLucia parameterization and the solution conditions described in the Methods section of the main text.
In this section, we consider a two-state model (i.e.~bonded or not bonded) for each domain and examine the simple case where pairs of strands hybridize to form dimers, but not larger multimers.
We denote the hybridization free energy between complementary domains on a pair of strands $i$ and $j$ by ${\upDelta G_{ij}}$.
The equilibrium probability that a strand $i$ is correctly hybridized with its putative neighbor strand $j$ is 
\begin{equation}
  p_{ij}(T) = \frac{\rho \exp\mleft(-\upDelta G_{ij}' / k_\text{B}T\mright)}
  {1 + \rho \exp\mleft(-\upDelta G_{ij}' / k_\text{B}T\mright)},
\end{equation}
where $\rho$ is the dimensionless strand number density, $k_\text{B}$ is the Boltzmann constant, $T$ is the absolute temperature, and we assume that all species are present in equal concentrations.
The hybridization free energies are written as ${\upDelta G_{ij}'}$ to indicate that we use the longest complementary subsequence of strands $i$ and $j$, which, due to the random sequence design, is occasionally longer than the intended domain length.
To calculate the total change in base-pairing during an annealing protocol (\figrefsub{fig-hybrid}{b}), we took the temperature derivative of the ensemble average of correctly formed base pairs,
\begin{equation}
  \text{Hybridization} = -\frac{\der}{\der T} \!\!\sum_{\substack{i<j\\j\in\mathcal{E}(i)}}\!\! l_{ij} p_{ij}(T),
\end{equation}
where $l_{ij}$ is the length of each hybridizing domain.

\section{Cluster population ratios}\label{subsec-cluster-population-ratios}
We assume that annealing is slow, so that nucleation is always rate-limiting.
We can write the nucleation barrier height as
\begin{equation}
  F^\dagger = -n^\dagger k_\text{B} T \ln \rhom + \varepsilon(T) E^\dagger + C,
\end{equation}
where $n^\dagger$ is the number of independent subunits in the critical nucleus, $E^\dagger$ is the number of 8-bp bonds in the critical nucleus, and $C$ is a constant that accounts for the (effective) number of parallel nucleation pathways, as well as the rotational entropy terms.
The bond energy $\varepsilon$ is a decreasing function of temperature, while the per-species monomer concentration $\rhom$, indicating the number of monomers per unit volume, also decreases as the reaction progresses.
Initially, we have $\rhoT$ of each species.
For simplicity, let us assume that, given this initial monomer concentration, the barrier is infinitely high above some critical temperature $T_0$.
Nucleation begins once ${T \le T_0}$, where $F^\dagger$ is finite.
(In reality, nucleation can begin as soon as the target structure, or any large cluster, becomes thermodynamically stable.  However, the nucleation rate is proportional to ${\exp(-F^\dagger/k_\text{B}T)}$, so the highest barrier that can be crossed depends on the cooling rate.)

Nucleation will proceed at a given temperature until $\rhom$ decreases to a point where $F^\dagger$ is again insurmountable.
Denoting this critical barrier height by $F_0^\dagger$, we can relate the final monomer concentration $\rhom$ at any temperature to the initial concentration at the critical temperature,
\begin{align}
  F^\dagger_0 - C &= -n^\dagger k_\text{B}T \ln \rhom(T) + \varepsilon(T) E^\dagger \\
  &= -n^\dagger k_\text{B}T_0 \ln \rhoT + \varepsilon(T_0) E^\dagger,
\end{align}
so that
\begin{equation}
  \frac{\rhoT}{\rhom} = \exp\mleft\{-\frac{E^\dagger}{n^\dagger} \left[\frac{\varepsilon(T)}{k_\text{B}T} - \frac{\varepsilon(T_0)}{k_\text{B}T_0}\right] \mright\}.
\end{equation}
Assuming that the intensity of each peak is proportional to the concentration of unassembled strands (m) or assembled structures (c), respectively, the ratio of the scattering intensities is
\begin{align}
  \frac{I_{\text{c}}}{I_{\text{m}}} &= \frac{R_{\text{h,c}}^6}{N R_{\text{h,m}}^6} \left(\frac{\rhoT - \rhom}{\rhom}\right) \\
  &= \frac{R_{\text{h,c}}^6}{N R_{\text{h,m}}^6} \left(\exp\mleft\{-\frac{E^\dagger}{n^\dagger} \left[\frac{\varepsilon(T)}{k_\text{B}T} - \frac{\varepsilon(T_0)}{k_\text{B}T_0}\right] \mright\} - 1\right),
\end{align}
where $N$ is the number of distinct subunits in the target structure.
Furthermore, because $\varepsilon/k_\text{B}T$ is a nearly linear function of $T$ in the range of interest (\figref{fig-cuboid-hybridisation-energies}), we expect the intensity ratio to have a functional form
\begin{equation}
  \frac{I_{\text{c}}}{I_{\text{m}}} = \text{const} \times \big\{\!\exp\mleft[-a (T - T_0)\mright]-1\big\},
\end{equation}
where $a = (E^\dagger/n^\dagger)(\der\beta \varepsilon/\der T)$ and $\beta=1/k_\text{B}T$.
Using a linear fit to the energies as a function of temperature at temperatures of interest (\figref{fig-cuboid-hybridisation-energies}), $\der\beta \varepsilon/\der T \approx \SI{0.34}{\per\kelvin}$.
From the theoretical free-energy profiles shown in the main text, we know that for edge BBs, $E^\dagger/n^\dagger=7/6$, whilst for face BBs, the ratio is $6/5$.
Hence we can estimate that $1/a\approx \SI{2.5}{\kelvin}$.

To calculate the intensity associated with each peak in the DLS data, we first fitted a sum of Gaussians to the distribution function, ${f(R_{\text{h}})}$.
We then numerically integrated the peak associated with the Gaussian function $f_\text{g}$, according to
\begin{equation}
  I_{\text{c/m}} = \int_0^\infty \min\mleft[f_{\text{g},\text{c/m}}(R_\text{h}),\, f(R_\text{h})\mright] \der\ln R_\text{h}.
\end{equation}
In reality, the appearance of aggregates at low temperatures, which tend to increase the mean $\Rh$ of the assembled population, means that the ratio of the scattering intensities is not exactly proportional to the ratio of the cluster concentrations.
However, this effect is relatively small over the range of temperatures of interest (approximately \SI{8}{kelvin} below $T_0$; see Fig.~4a).
Instead, the exponential increase in the intensity ratio as a function of decreasing temperature shown in Fig.~5c is driven primarily by an exponential decrease in the scattering intensity of the unassembled population upon cooling below $T_0$.
Such behavior is consistent with the theoretically predicted evolution of the unassembled-strand population shown in Fig.~5b.

\clearpage

\section{Supplemental figures}
\newpage$\,$

\onecolumngrid

\begin{figure}[h]
\includegraphics{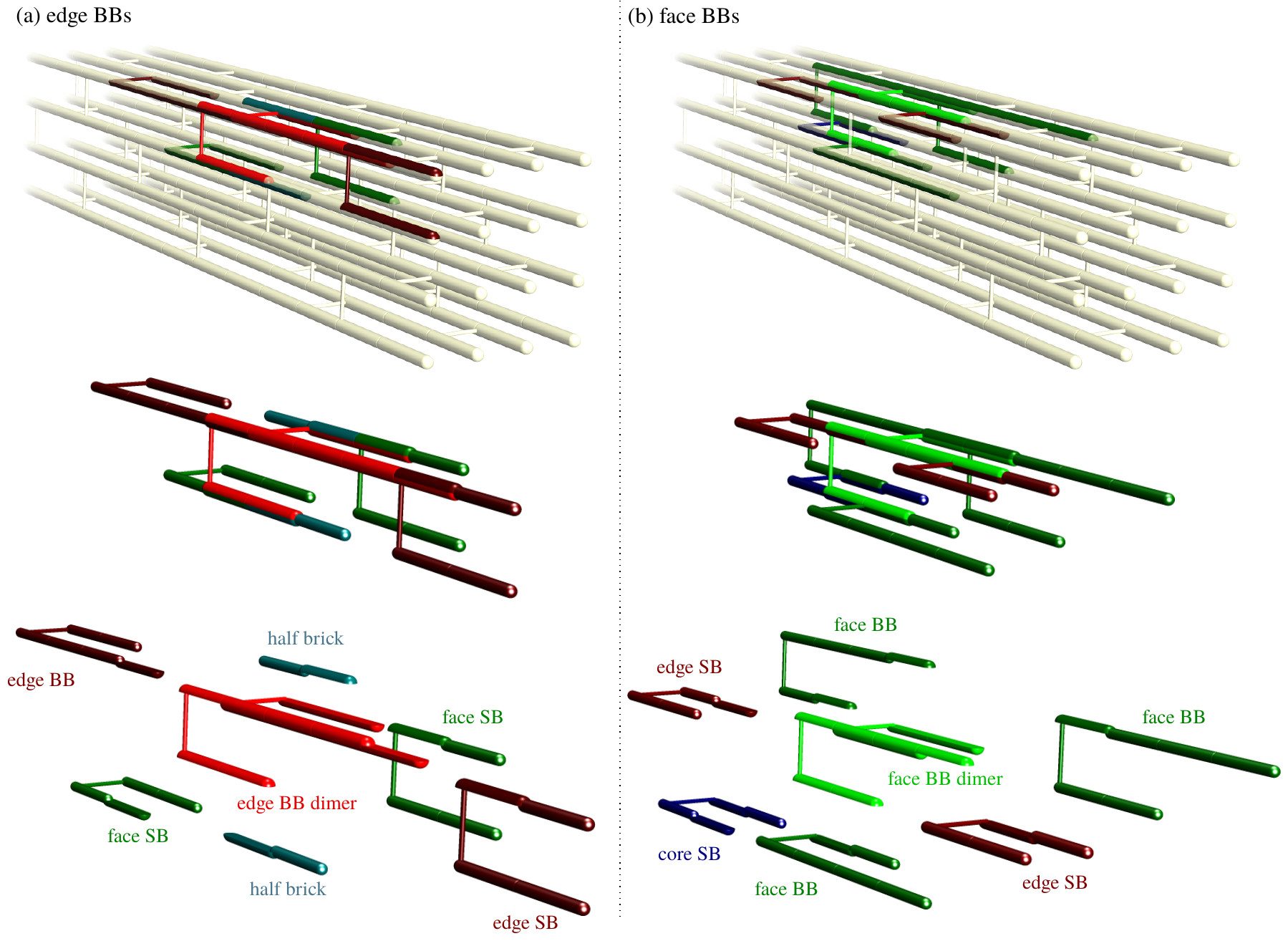}
\caption{Boundary brick dimers and their nearest neighbors in a schematic representation. In the top panel, the location within the target structure is shown. In the bottom panel, the neighbors are shown spread out and labeled to make their identification clearer.  `SB' stands for a 32-nt `scaffold brick'. The edge-BB system's nucleation properties were also investigated by merging some bricks, as described in the main text. In particular, the `merged-A' building block corresponds to the edge BB dimer shown in red. The `merged-B' building block corresponds to the edge BB dimer and one of the face SBs shown in dark green. Either one of these face SBs could have been chosen, as both of them have direct connections to core strands. In our simulations, the face SB that is merged with the edge BB dimer is the one whose center of mass is nearer the cuboid's principal axis in the target structure. }\label{fig-BB-illustrations}
\end{figure}

\clearpage
\twocolumngrid
 
\begin{figure}[h!]
  \includegraphics{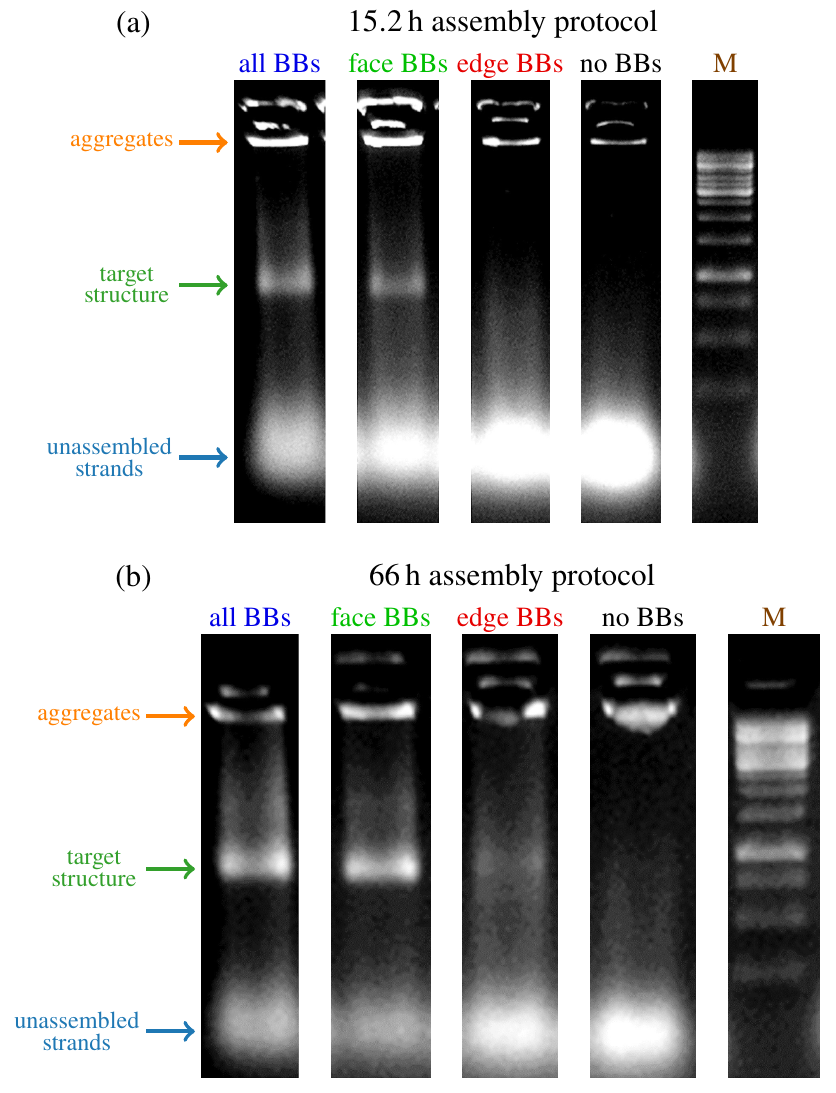}
\caption{Gel electrophoresis of samples at the end of (a) 15.2 hour or (b) 66 hour linear assembly protocols. Samples were assayed in \SI{2}{\percent} agarose gel. Lane M contains a GeneRuler \SI{1}{kb} ladder that was used to reference the assembly yield.  The bands corresponding to the target structures, unassembled strands and aggregates are indicated.}\label{fig-gels}  
\end{figure}

\vskip0.25in
\begin{figure}[h]
\includegraphics{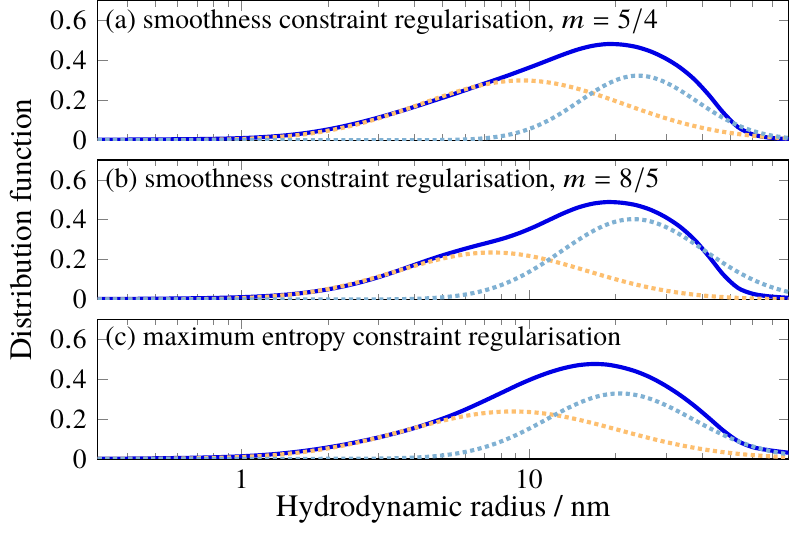}
\caption{Example size distribution functions (solid blue lines) for the all-BB system at \SI{310}{\kelvin} determined using three regularization methods.  We used a smoothness constraint functional~\cite{Provencher1982} with smoothing exponents~\cite{Hansen2017} of (a) $m=5/4$ and (b) $m=8/5$, as well as (c) a maximum entropy constraint with a Gaussian prior distribution~\cite{Hansen2017}.  Dotted lines show the Gaussian functions determined by fitting a linear combination of Gaussians to each distribution function.}\label{fig-DLS-regularisation}
\end{figure}

\newpage
\begin{figure}[h]
\includegraphics{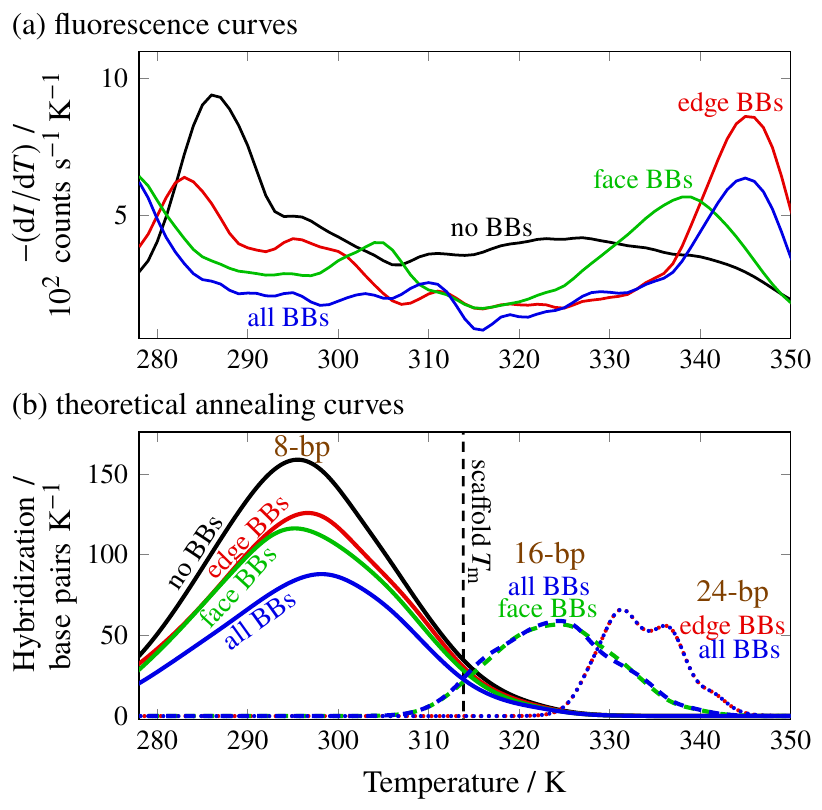}
\bigskip
\caption{(a) The derivative of the fluorescence signal $I$ with respect to the temperature obtained from a \SI{15.2}{\hour} annealing protocol. (b) The corresponding theoretical annealing curves (see \secref{subsec-hybridization}).  In agreement with the experimental fluorescence data, contributions from 16- (dashed lines) and 24-bp (dotted lines) hybridizations dominate at higher temperatures.  The predicted scaffold-strand \TM{} is also shown.}
\label{fig-hybrid}
\end{figure}

\vskip0.15in
\begin{figure}[h]
\includegraphics{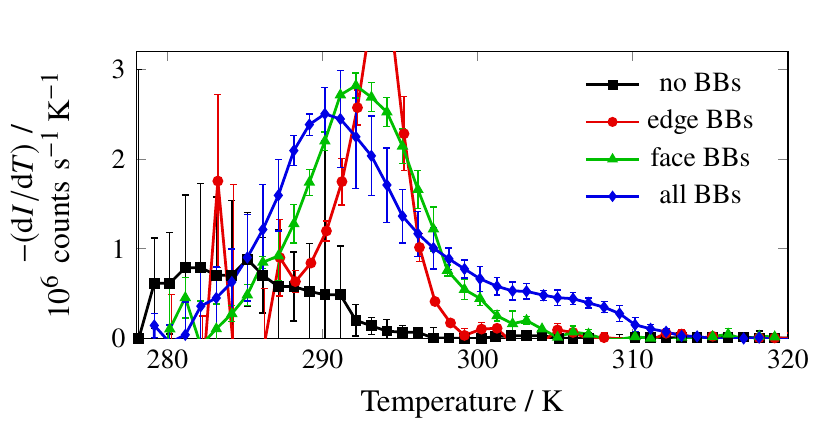}
\caption{The derivative of the static light scattering intensity $I$ with respect to temperature for self-assembly following the \SI{15.2}{\hour} annealing protocol.}\label{fig-cuboid-static}
\end{figure}

\clearpage
\onecolumngrid
\begin{samepage}
\begin{figure*}[h]
\includegraphics{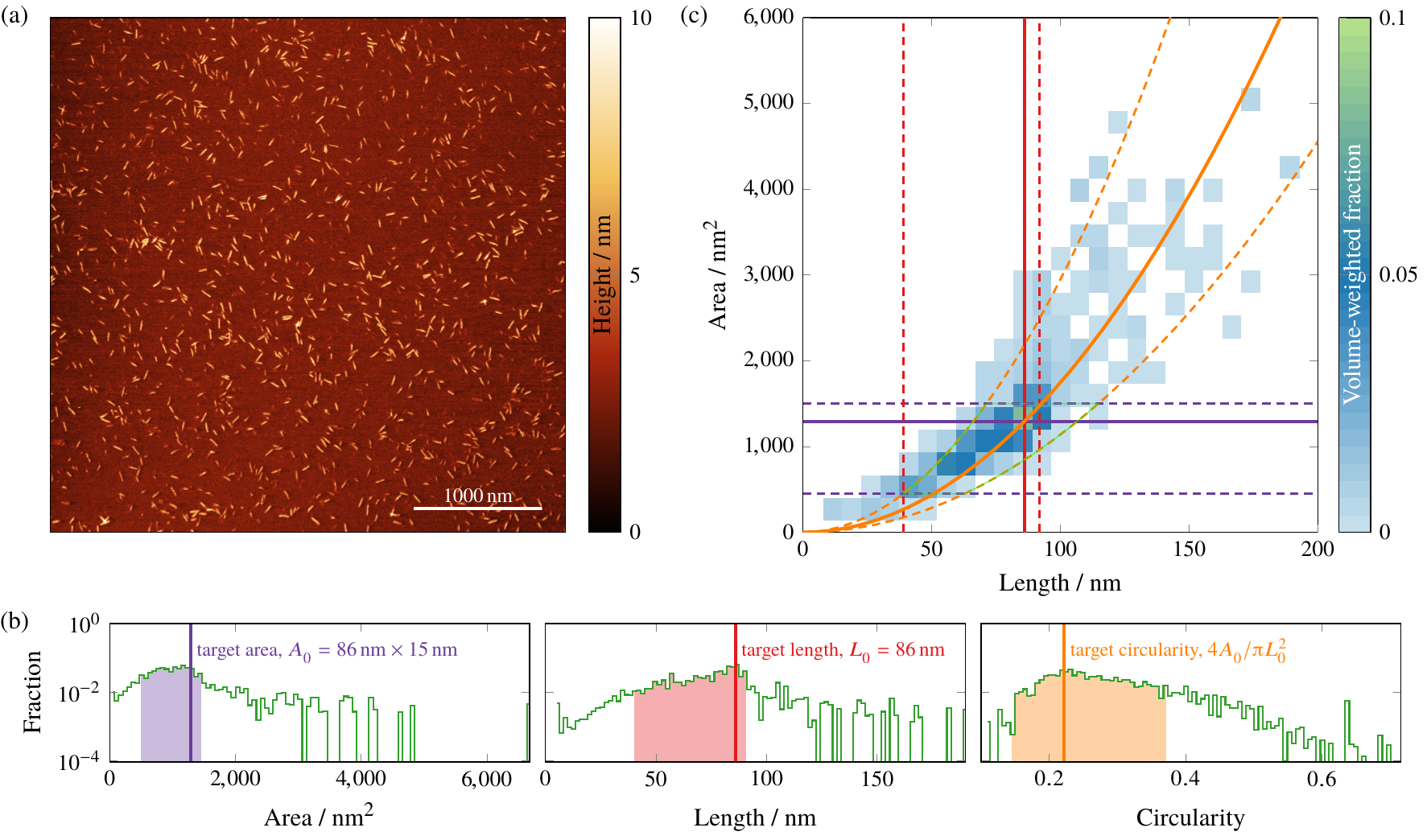}
\vskip-2ex
\caption{Criteria for identifying target structures via AFM.  (a) AFM image of the positive-control all-BB structures (see \secref{subsec-annealing}).  (b) Area, length and circularity distributions of the particles identified in this image (see Methods).  All distributions are weighted by the particle volume to prevent tiny particles from skewing the distributions.  The solid vertical lines show the expected values for an ideal target structure when treating the cuboid as a cylinder with diameter $d\sim\SI{15}{\nano\metre}$.  (c) Two-dimensional volume-weighted distribution of imaged particle areas and lengths.  The solid lines indicate the expected values for an ideal target structure, while the dashed lines correspond to the boundaries of the shaded regions in panel (b); the parabolic curves show the circularity, $4A/\uppi L^2$.  Notably, the peak of this distribution (green square) coincides with the expected area, length and circularity.  Based on this distribution, we chose to use the area and circularity criteria ($\SI{450}{\nano\metre^2} \le A \le \SI{1500}{\nano\metre^2}$ and ${0.145 \le 4A/\uppi L^2 \le 0.375}$, indicated by the translucent green lines) to identify particles as correctly assembled structures.  Using these criteria, the AFM-determined yield of the positive control is \SI{53}{\percent}.}\label{fig-AFM-all-BB-positive-control}
\end{figure*}

\begin{figure*}[h]
  \includegraphics{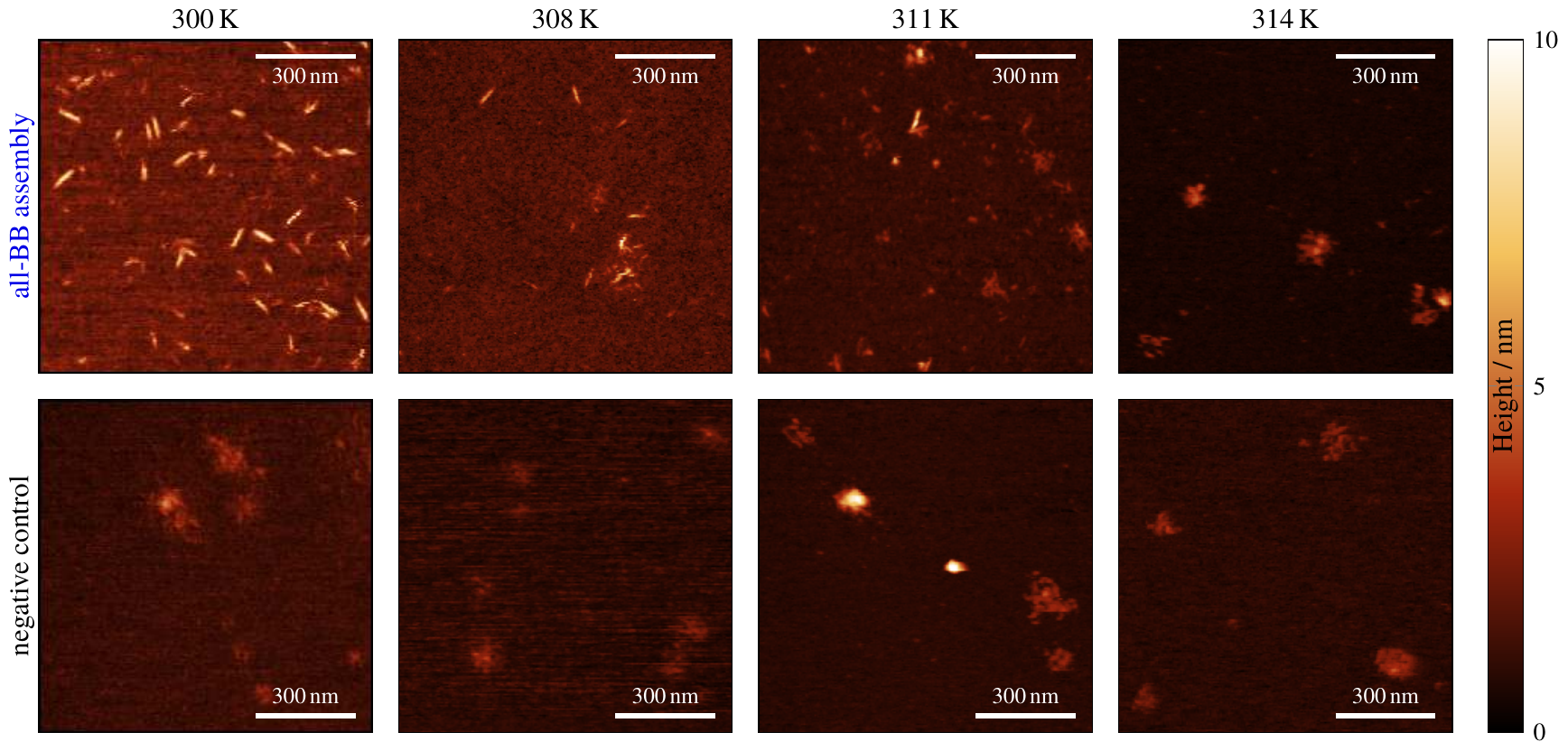}
\vspace{1ex}
\caption{AFM images for all-BB structures (top row) and negative controls (bottom row).  Samples were prepared by rapidly quenching the system from the indicated temperature (column labels) during the annealing protocol (see \secref{subsec-afm-protocols}).  The negative control consists of a collection of similar-length oligonucleotides that were not designed to have complementary sequences.  The bright amorphous clusters seen in the negative control and the all-BB structures quenched from high temperatures indicate large aggregates of oligonucleotides that form during quenching.}\label{fig-AFM-all-BB-images-distributions}
\end{figure*}
\end{samepage}

\clearpage
\twocolumngrid

\begin{figure}[h]
\includegraphics{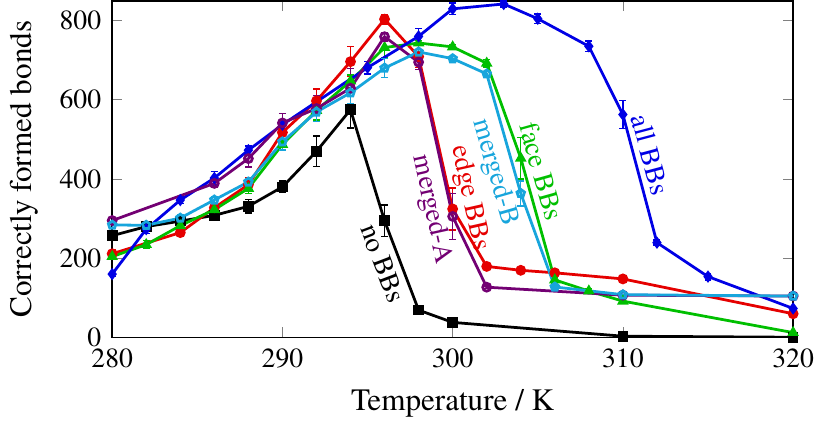}
\caption{The number of correctly formed bonds in the system as a function of temperature from Monte Carlo simulations. Each data point corresponds to an average over ten independent simulations in the long time limit once nucleation has occurred. Error bars give the standard deviation in each case.}\label{fig-cuboid-lowConc-clustBonds}
\end{figure}

\vskip0.25in
\begin{figure}[h]
\includegraphics{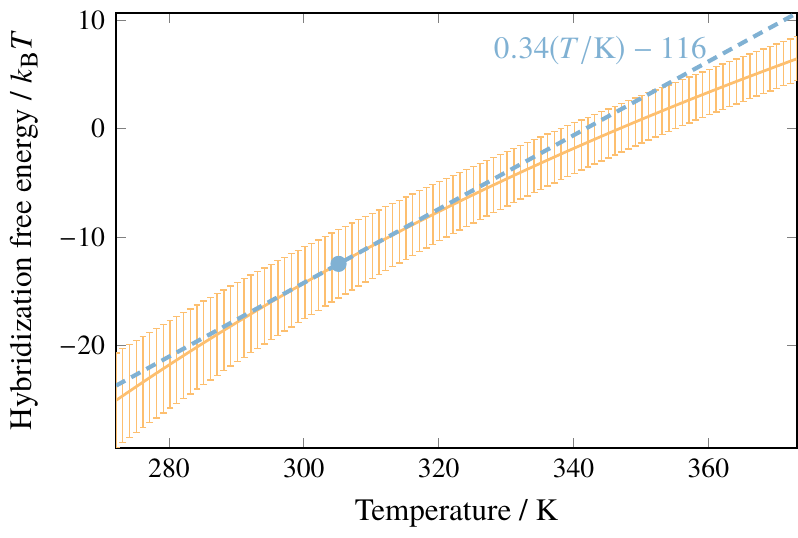}
\caption{The mean hybridization free energies of 8-bp interactions as a function of temperature for the no-BB structure, computed via the SantaLucia thermodynamic model~\cite{SantaLucia2004}, with error bars reflecting the standard deviation. The tangent to the curve at \SI{305}{\kelvin} is also shown, demonstrating that the hybridization free energy is well described by a linear function over the region of interest (\SIrange{295}{315}{\kelvin}).}\label{fig-cuboid-hybridisation-energies}
\end{figure}

\vskip0.25in
\begin{figure}[h]
\includegraphics{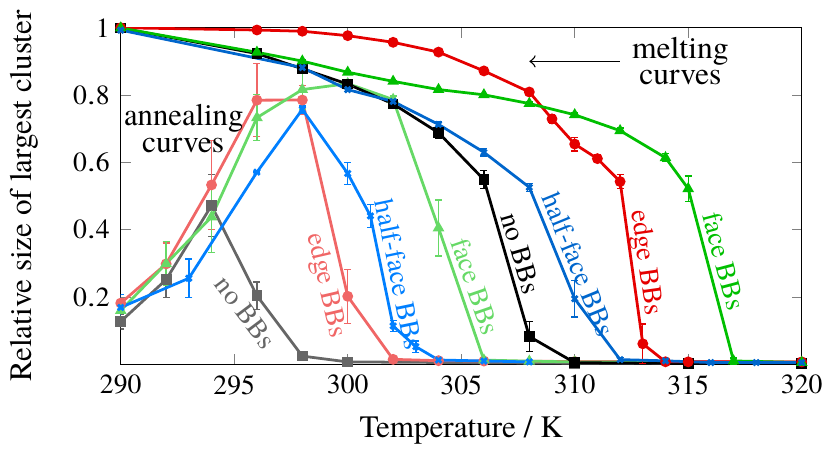}
\caption{Melting and annealing curves for the half-face BB structure in Monte Carlo simulations (cf.~Fig.~4c and Fig.~6b).}\label{fig-face-halfface} 
\end{figure}

\newpage

\begin{figure}[h]
\includegraphics{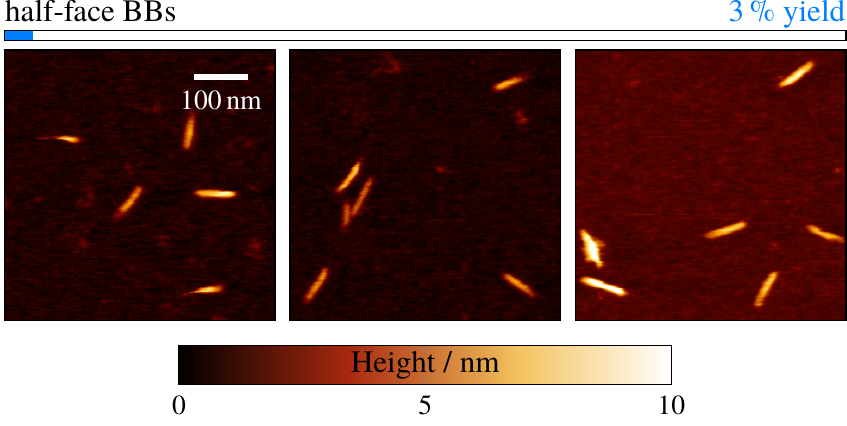}
\vskip0.25ex
\caption{AFM images and the yield, as determined by gel electrophoresis, for the half-face BB structure.}\label{fig-sample-afm-comb-zoomed-half}
\end{figure}

\vskip0.2in
\begin{figure}[h]
\includegraphics{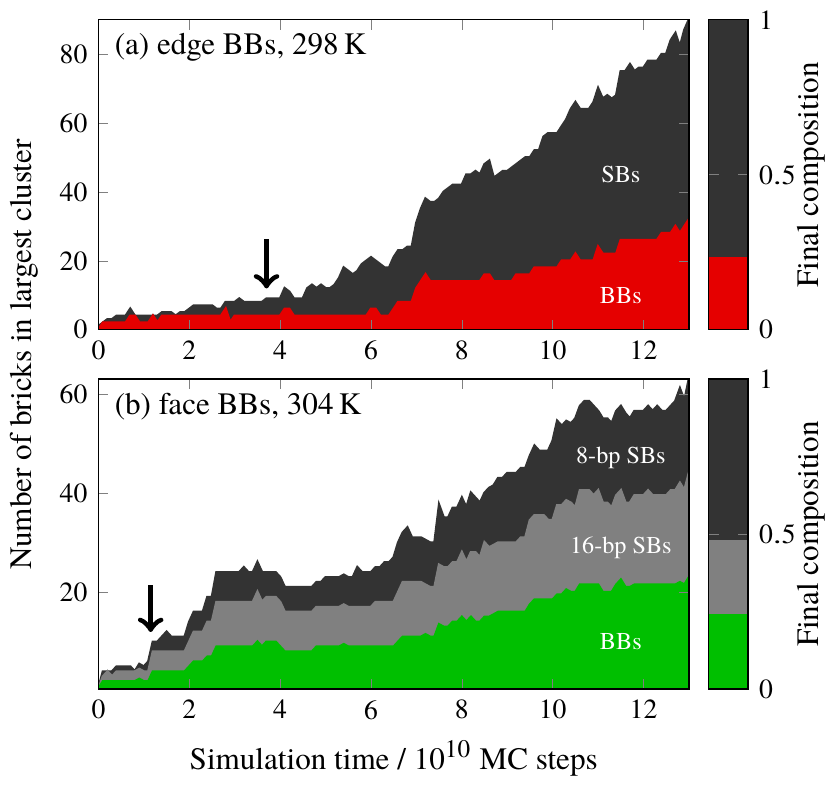}
\vskip0.25ex
\caption{The size of the largest cluster as a function of time in representative Monte Carlo trajectories for the (a)~edge-BB and (b)~face-BB systems.
  The composition of the largest cluster is shown in terms of boundary bricks (BBs) and scaffold bricks (SBs).
In the face-BB system, the scaffold bricks are further divided into those that have only 8-bp hybridizations with neighboring bricks and those which form 16-bp hybridizations with face BBs.
The relative proportions of each type of brick in the target structure are shown in the right-hand panel in each case.
Both trajectories start at roughly the point where a fluctuation leads to nucleation and subsequent growth.
Arrows indicate approximately where the theoretically predicted post-critical nucleus (see Fig.~7) first appears in each trajectory.
In (a), this cluster has two BB dimers (four BBs) and four SBs.
In (b), this cluster has three BB-SB dimers (three BBs and three 16-bp SBs) and two 8-bp SBs.
In both cases, the cluster size increases rapidly after this nucleation event.
BBs dominate in the early part of the growth phase shown here, where they comprise a greater proportion of the post-critical clusters than they do in the final assembled structure.
}\label{fig-sim-pathways}
\end{figure}

\clearpage
\onecolumngrid

\begin{figure}[h]
\includegraphics{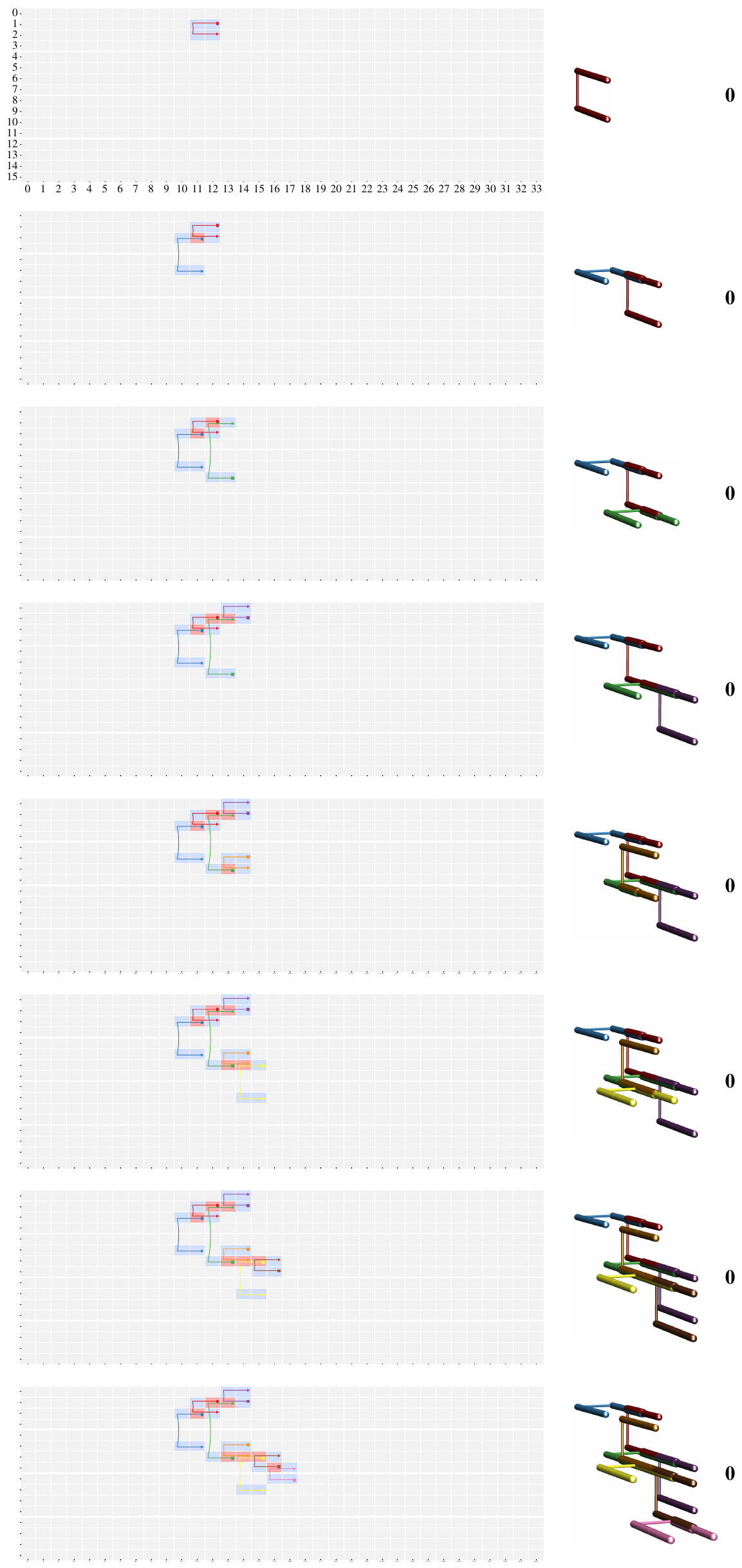}
\caption{Example low free-energy nucleation pathway for the no-BB system in two representations: a Cadnano-style connectivity diagram and a three-dimensional rendering. In the latter, DNA brick domains are represented by cylinders. Non-bonded domains are represented by smaller cylinders, while where two DNA bricks are bonded, larger multicolored cylinders are used. Each new monomer or multimer added to the cluster along the nucleation pathway is colored in a different hue. The bold number to the right of each structure indicates the number of multimers in the structure.}\label{fig-minfe-pathways-noBB}
\end{figure}

\clearpage
\begin{figure}[h]
\includegraphics{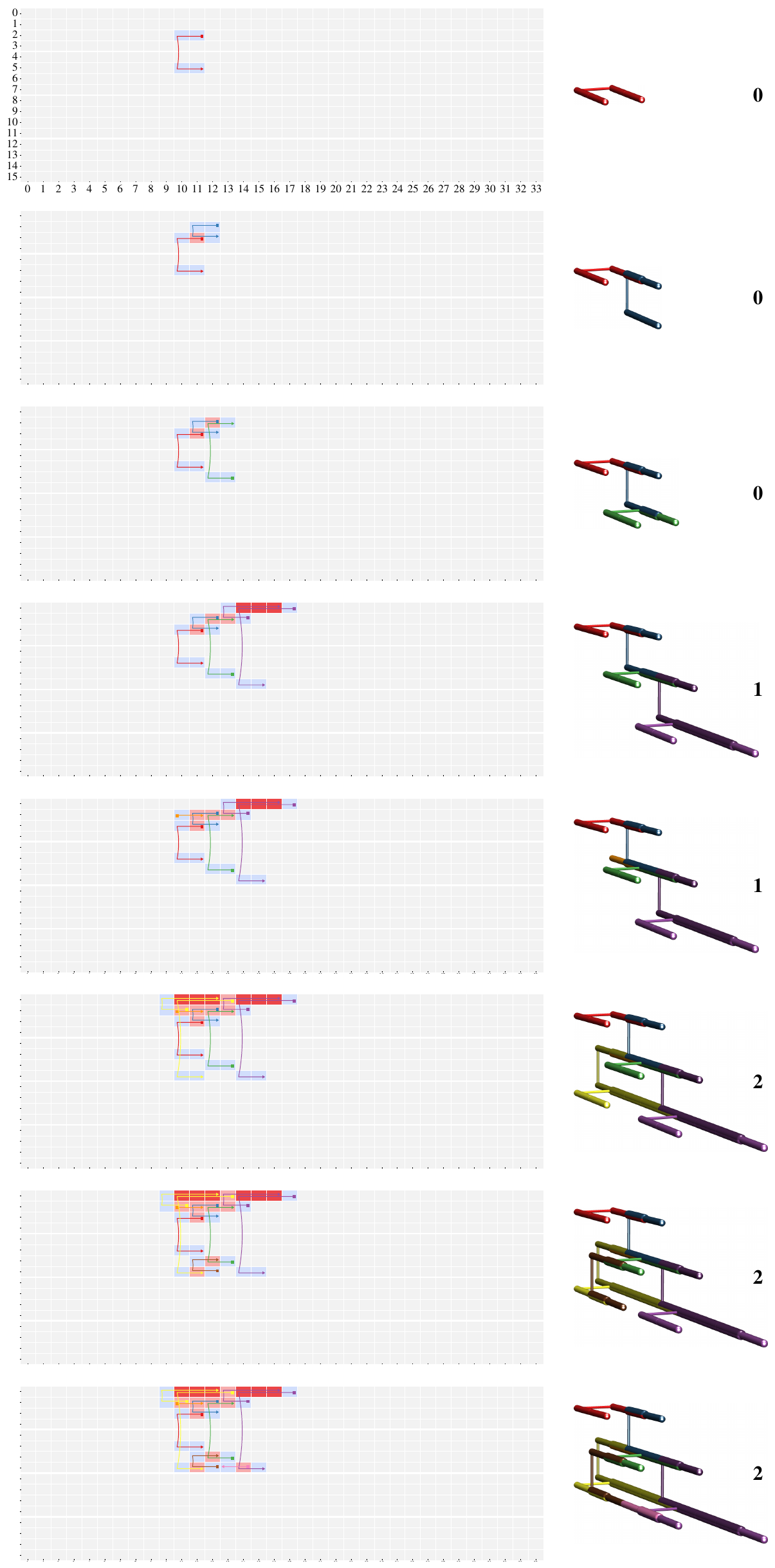}
\caption{Example low free-energy nucleation pathway for the edge-BB system in two representations, as in Fig.~\ref{fig-minfe-pathways-noBB}.}\label{fig-minfe-pathways-edgeBB}
\end{figure}

\clearpage
\begin{figure}[h]
\includegraphics{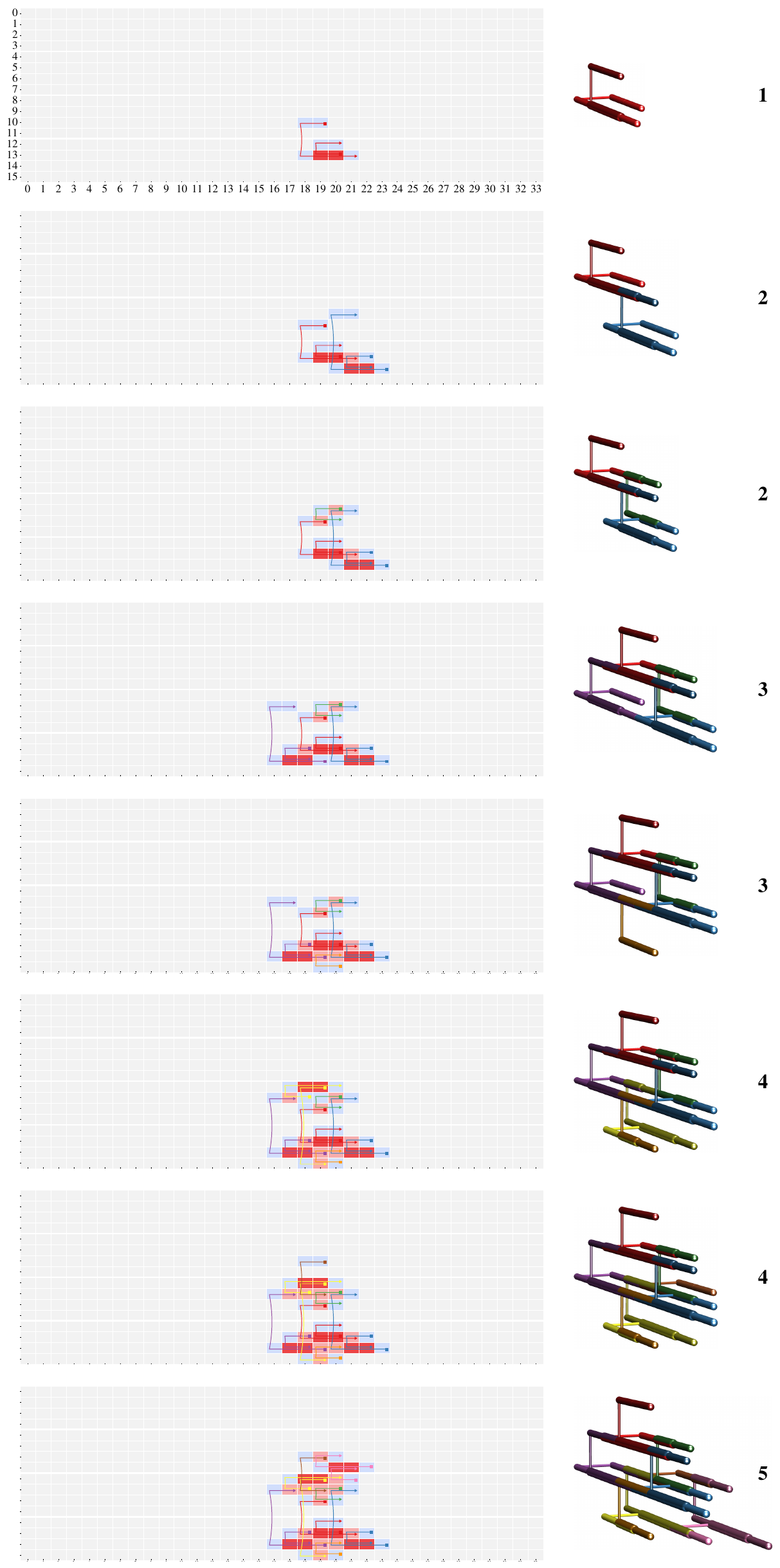}
\caption{Example low free-energy nucleation pathway for the face-BB system in two representations, as in Fig.~\ref{fig-minfe-pathways-noBB}.}\label{fig-minfe-pathways-faceBB}
\end{figure}

\clearpage
\twocolumngrid

\section{DNA brick sequences}\label{sec-seqs}

The following sequences comprise our library of DNA bricks.
\begin{center} \scriptsize\ttfamily
\xentrystretch{-0.125}
\tablefirsthead{\toprule \rmfamily\textbf{ID}& \rmfamily\textbf{Sequence}\\ \midrule}
\tablehead{
\toprule \rmfamily\textbf{ID}& \rmfamily\textbf{Sequence} & ~ & ~ & ~ & ~ & ~\\ \midrule}
\tabletail{\midrule \multicolumn{7}{r}{\rmfamily\textit{Table continues ...}} \\ \midrule}
\tablelasttail{\\ \bottomrule}
\begin{xtabular}{>{\rmfamily}p{0.4cm}p{1.05cm}p{1.05cm}p{1.05cm}p{1.05cm}p{1.05cm}p{1.05cm}}
1 & TTCTTAAA & TTTTTTTT\\ 
2 & TTGGCTGA & TTTTTTTT\\ 
3 & TTTTTTTT & CCGCGTAA\\ 
4 & TTTTTTTT & CCAACAGG\\ 
5 & TATGGTGA & TTTTTTTT & TTTTTTTT & TGGCACCC\\ 
6 & ATCCGAAG & TTTTTTTT & TTTTTTTT & CGCGGACA\\ 
7 & CGCTGATC & TTTTTTTT & TTTTTTTT & GGGGATGA & ATCCTTCC & AACATCTC\\ 
8 & TCTGATAT & TTTTTTTT & TTTTTTTT & GCACTGCC & CCTACTCT & CACCTTTC\\ 
9 & ATATCAGA & GAGCACAG\\ 
10 & TCCGGGAT & GGGTGCCA\\ 
11 & AGAGTCTG & GAGATGAT\\ 
12 & GCTTCGCG & CCTGTTGG\\ 
13 & TAATATAA & TGTCCGCG & GATCAGCG & AATTCGAC\\ 
14 & AGTTAGCA & TTACGCGG & GTACCCTG & GACCGCTA\\ 
15 & GGAAGGAT & TCATCCCC & TTTAAGAA & TGATCGCA & CCAGGTTA & AGTGGCTC\\ 
16 & TACATCTC & GCTCGTCC & TCACCATA & ATCCACCA & CCCATGCA & GAAAGATA\\ 
17 & GTGCGCTT & CTGTGCTC & ATCCCTTG & CTCCCGAA\\ 
18 & TCCCTAAA & GTCGAATT & GAGATGTA & AGTTTCAC\\ 
19 & ACCTTTTA & ATCATCTC & CAGACTCT & TTTTTTTT & TTTTTTTT & GAGCCGAT\\ 
20 & GGGAGATT & TAGCGGTC & CAGGGTAC & TTTTTTTT & TTTTTTTT & GGACGAGC\\ 
21 & CAGTCTTT & GAGATGTT & GAACATCT & TTTGGTTG\\ 
22 & AAGGTAAT & GTGAAACT & CAATGATA & TGACGGAT\\ 
23 & ACATCTCG & GAAAGGTG & AGAGTAGG & GGCAGTGC & TCAGCCAA & ACAGTGGG\\ 
24 & CTTACACT & TTCGGGAG & CAAGGGAT & ATCGGCTC & CTTCGGAT & TCTCTGGC\\ 
25 & TATCATTG & GCCAGAGA & TGCTAACT & GTATGACG & GACCAGGT & CTCCTAGG\\ 
26 & TGCATGGG & TGGTGGAT & CGCGAAGC & TGTCGTGA & ATGATCGG & TCATATGT\\ 
27 & ACAATGGT & ATCTTTTG & AAGCGCAC & GTAAACCG & TAATTCGG & AGCCCGGC\\ 
28 & ACCTGGTC & CGTCATAC & TAAAAGGT & AGAAGTAA & GTAGGGTA & CCGCTGGG\\ 
29 & GCTCTTAA & TATCTTTC & GGTTGCGG & GAGTAAGG\\ 
30 & GTTTAGTG & ATCCGTCA & ACCATTGT & TATTGGCC\\ 
31 & CACCAGAA & GAGCCACT & TAACCTGG & TGCGATCA & ATCCCGGA & CACGTAAA\\ 
32 & CCACAAAG & TCTTGTAT & AAAGACTG & TGCGAGAT & CTTTACGC & GCTTGAAC\\ 
33 & TTCCTTTT & CAACCAAA & AGATGTTC & CCCACTGT & TTATATTA & CAAAAGAT\\ 
34 & CCGAATTA & CGGTTTAC & CGAGATGT & TTACTTGC & ACTTGGGA & TTAGGATC\\ 
35 & CTTGCCTG & GGCCAATA & CTTTGTGG & TCCACTAT\\ 
36 & CAGTCCGA & CCTAGGAG & CGGTCTGG & CTGTGAGG\\ 
37 & GCGTAAAG & ATCTCGCA & TTCTGGTG & CCAAAACT & AATCATAC & TCACGGTT\\ 
38 & TTCAGCAG & CCTTACTC & CCGCAACC & TTTACGTG & TTTAGGGA & ATACAAGA\\ 
39 & GTGCTCTT & GAGCGGTG & TTAAGAGC & CTCGCTTC & AAGTCCGT & GTCGGTAG\\ 
40 & AAAACTTC & ACATATGA & CCGATCAT & TCACGACA & AATCTCCC & GCTAGATG\\ 
41 & GGGAGCTG & GCCGGGCT & ATTACTTT & TGCCATAT\\ 
42 & GTTGTTCC & ATAGTGGA & AAGAGCAC & GTTAACTC\\ 
43 & GTAGTTTA & CCCAGCGG & TACCCTAC & TTACTTCT & AGTGTAAG & CAACCCAT\\ 
44 & CGACTCAA & CCTCACAG & CCAGACCG & CATCTAGC & ATTACCTT & CACCGCTC\\ 
45 & GCCGACTC & GTTCAAGC & ACTATTAT & CTGGCTAT\\ 
46 & ACTGAGGC & GAGTTAAC & TTATGTTC & TAACCTGC\\ 
47 & TCATGTGG & GATCCTAA & TCCCAAGT & GCAAGTAA & AAAAGGAA & TTATGAGA\\ 
48 & GATCACAA & ATATGGCA & AAAGTAAT & ATGGGTTG & CACTAAAC & TATAGCAA\\ 
49 & GAACATAA & TTGCTATA & TCGGACTG & TGGGCGTA & CCTACACC & ACAGATGC\\ 
50 & ACGGACTT & GAAGCGAG & GAAGTTTT & GCGACTCA & TTGCTGAG & TTCGACGC\\ 
51 & ATATAGGT & CTGACACC & CAGCTCCC & AGGTAAGT & GTATTATC & GCGTCAAT\\ 
52 & GGTGTAGG & TACGCCCA & TAAACTAC & TTTTCCAC & GTTGCCCT & CGACCGAT\\ 
53 & GGAGGGTT & CTACCGAC & ATATTGCT & TTGCCCTT\\ 
54 & TGTTGAGG & GCAGGTTA & ACCTATAT & AGCCATTT\\ 
55 & AGCTTGGG & AACCGTGA & GTATGATT & AGTTTTGG & CTGCTGAA & TGTCCGTG\\ 
56 & CGACCATT & TGACACCC & GAGTCGGC & CTGACAGA & CATCTTTT & GGGGTCGT\\ 
57 & AGCCTACG & ATAGCCAG & ATAATAGT & TCTCATAA & CAGGCAAG & GGTGTCAG\\ 
58 & GATAATAC & ACTTACCT & CCACATGA & AGGTCTAC & CAGAAACT & ATGTACCG\\ 
59 & GCGGAGAG & AAATGGCT & AATGGTCG & ACAATATG\\ 
60 & CGTCCAGC & GCATCTGT & GATCCACG & TCAGATTT\\ 
61 & AAAAGATG & TCTGTCAG & CCCAAGCT & TAGCATAA & AACTGTTG & TGTCCTAT\\ 
62 & GTCCAAGT & AAGGGCAA & AGCAATAT & CACGGACA & GGAACAAC & GGGTGTCA\\ 
63 & GCGTATGC & TGAAATGG & AACCCTCC & TACGTTTT & GCGATGGT & GATGGGAA\\ 
64 & TTTAGTTT & GCGTCGAA & CTCAGCAA & TGAGTCGC & TTGAGTCG & GTGGATAA\\ 
65 & ATCGCTCA & ATTGACGC & GACTTTGT & GTAATCTC\\ 
66 & CATACACA & CATATTGT & GCATACGC & CCTGGTTT\\ 
67 & CTGAGCGT & ATCGGTCG & AGGGCAAC & GTGGAAAA & TTGTGATC & CCCTCCCA\\ 
68 & AATAGGCT & ATTCAAGA & GCTGGACG & TGCGCGAG & GCACGTAT & GTCTTGAG\\ 
69 & TAGGGCTA & AAATCTGA & CGTGGATC & TTATCCAC & GCCTCAGT & CCATTTCA\\ 
70 & ACCATCGC & AAAACGTA & AAACTAAA & TGACGTTT & GAGGTGAT & ACGTAAAG\\ 
71 & GCGGCTGG & ACGACCCC & GCCGTGCG & ACGACTGA\\ 
72 & GTAATGCG & AAACCAGG & AGCCTATT & CATCAAGG\\ 
73 & GATATCCA & CGGTACAT & AGTTTCTG & GTAGACCT & CGTAGGCT & AGAGCGGT\\ 
74 & TATTAAGT & TTTCGCCG & TGAGCGAT & CAATGACC & TCATATGG & ACACACCA\\ 
75 & GATCCTGA & GAGATTAC & ACAAAGTC & TGGGAGGG & CCTCAACA & TCTTGAAT\\ 
76 & ATACGTGC & CTCGCGCA & ACGCTCAG & CCGGTCTG & TAGGGTGA & CGTATTAG\\ 
77 & CAAATCCC & TTCCCATC & AAATACGA & CTAAGCCG\\ 
78 & CGTTGCAC & CCTTGATG & ACTTAATA & TTCAGGCT\\ 
79 & CGTTTGTC & ATAGGACA & CAACAGTT & TTATGCTA & ACTTGGAC & CGGGCATC\\ 
80 & GCAGGGTG & AAAGTGAG & CCAGCCGC & GCGCAGGG & CTCGACGG & TTAATTTG\\ 
81 & ACTCTCGT & TCAGTCGT & CGCACGGC & ACCGCTCT & CTCTCCGC & CGGCGAAA\\ 
82 & CCATATGA & GGTCATTG & TGGATATC & TCCGTGCA & GTAAACGA & TAGGTTTC\\ 
83 & CCAGTAAT & AGCCTGAA & CACCCTGC & ATCGACGG\\ 
84 & GCTAGCAT & CTCAAGAC & CTGGGAAT & TAGGTAAG\\ 
85 & CCGTCGAG & CCCTGCGC & GACAAACG & CATATCCT & GTTTACTT & AGGGATGG\\ 
86 & TCTGGCGC & CGGCTTAG & TCGTATTT & GATGCCCG & TGTGTATG & CTCACTTT\\ 
87 & GAAAGATC & ATAACGTT & GGGATTTG & TTCCGCCT & ATAGGTAA & CTGGGTTT\\ 
88 & GTCAATTT & CTTTACGT & ATCACCTC & AAACGTCA & TAGCCCTA & CGGATCAA\\ 
89 & GAATTATA & TGGTGTGT & CCATATCA & ACAAACCC\\ 
90 & GCGGCGAA & CCGTCGAT & GATCTTTC & GGACCGAG\\ 
91 & GGCTTGCA & CTAATACG & TCACCCTA & CAGACCGG & TCAGGATC & ATTTTGGA\\ 
92 & TTTGTAAA & CTTACCTA & ATTCCCAG & TTGATCCG & CGCATTAC & AACGTTAT\\ 
93 & CGGCCCGC & CAAATTAA & TAGTCGAT & TACGCTTC\\ 
94 & AGCTCATT & CTCGGTCC & AATTCGAG & CATGTCCC\\ 
95 & AAGCAAGA & GAAACCTA & TCGTTTAC & TGCACGGA & ACGAGAGT & GCTGGCAT\\ 
96 & TGGGCCCG & GGGTTTGT & TGATATGG & TCCAAAAT & GTGCAACG & GGAAACAC\\ 
97 & CTCGAATT & GTGTTTCC & ATGCTAGC & GCCACACG & ACTACTTA & AGACTTAG\\ 
98 & TTACCTAT & AGGCGGAA & AAATTGAC & ATTTCTAC & AATCGTGA & CGGTTTCT\\ 
99 & CTTCTACC & TCTTAATG & TATAATTC & ATCCTGTG & GATTACGG & CGGCTTAG\\ 
100 & TAAGTAGT & CGTGTGGC & TGCAAGCC & CTCACCTG & ACCTCTTA & GCACAATA\\ 
101 & CCTAACAT & AAACCCAG & CGTGACGG & TGAATCTT\\ 
102 & GTGTAAGA & GGGACATG & GGTAGAAG & GAACTCTG\\ 
103 & ATTACCGT & CCATCCCT & AAGTAAAC & AGGATATG & GCGCCAGA & CCACAATG\\ 
104 & GCCGATGG & AAAACGAC & GCGGGCCG & GGAAGGAC & CTTTAGGC & CCAAGGAG\\ 
105 & TCGCAACG & GAAGCGTA & ATCGACTA & ATGCCAGC & ATTACTGG & CATTAAGA\\ 
106 & CCGTAATC & CACAGGAT & TCTTGCTT & ATCTTAGT & CCCCAAGT & GCTACATG\\ 
107 & CATGGAAG & CAGAGTTC & CCATCGGC & AAAATTGA\\ 
108 & AACCGCGC & CTAAGTCT & TTAGAATC & ACCTTAGG\\ 
109 & GCCTAAAG & GTCCTTCC & ACGGTAAT & CGGGGGTG & TACGAACT & CAACGCAC\\ 
110 & CTCAGTTT & AAGATTCA & CCGTCACG & CATTGTGG & TTCGCCGC & GTCGTTTT\\ 
111 & CGACCTCT & AGATCCCA & ATGTTAGG & AGCTAACA & GATATTTC & AAGGACAC\\ 
112 & TCTTCAAA & AGAAACCG & TCACGATT & GTAGAAAT & TTTACAAA & GACGGACA\\ 
113 & AATGTTCT & CTAAGCCG & AGTAAACA & ATCAAGGA\\ 
114 & GAAGAGGT & TCAATTTT & AGAGGTCG & AGGCGAGT\\ 
115 & GTAGGCTC & TATTGTGC & TAAGAGGT & CAGGTGAG & CGGGCCCA & ATTGTACC\\ 
116 & ACGGAAAC & TCCTACCC & GCGCGGTT & GATGTGCC & TGCGAAAA & GCAAGCCG\\ 
117 & GTGGCGGA & CCTAAGGT & GATTCTAA & TGTCCGTC & AATGAGCT & TGGGATCT\\ 
118 & GAAATATC & TGTTAGCT & TTTGAAGA & AGGCTCCG & TCATTCCG & TCCCCCAA\\ 
119 & ACTCCGGC & CTCCTTGG & TATATTAG & CTTACCCA\\ 
120 & GACCCGAC & ACTCGCCT & GTTTCCGT & GCGTTCTT\\ 
121 & GCGATTCG & CATGTAGC & ACTTGGGG & ACTAAGAT & CGTTGCGA & ATTTTCGC\\ 
122 & GTTACCGC & TATAATTA & AGAACATT & CGGGTTCG & GGGAATTG & AGTACAAG\\ 
123 & GCGCAGTT & TCCTTGAT & TGTTTACT & GGTACAAT & TCTTACAC & GGGTAGGA\\ 
124 & TTTTCGCA & GGCACATC & GAGCCTAC & CCGTCTAT & AGCAGAAT & GGTATAAA\\ 
125 & CACCTACC & GTGTCCTT & CGTCCTGT & CCCGTTCC\\ 
126 & ATATTACA & AAGAACGC & GCGGTAAC & ATCACATA\\ 
127 & CGAGTCGT & GTGCGTTG & AGTTCGTA & CACCCCCG & AAACTGAG & GAGAACAT\\ 
128 & CCTAGGCG & TAAGACCT & GCCGGAGT & AGACGTGC & CACGCGAG & CTATGTAG\\ 
129 & CGTCGTCT & TGGGTAAG & CTAATATA & GCGAAAAT & CTTCCATG & TAATTATA\\ 
130 & CAATTCCC & CGAACCCG & CGAATCGC & AGGACGAT & CTCGTTCG & GGTGCTAC\\ 
131 & ACGTGGTC & TATGTGAT & CGCCTAGG & CTAACCTC\\ 
132 & CGGACACA & CGGCTTGC & TGCGACCA & AGTTTTAT\\ 
133 & CTCGCGTG & GCACGTCT & ACGACTCG & TCGAACGA & GCTCGGGT & AGTCTCAA\\ 
134 & ATTTTGGC & GGAACGGG & ACAGGACG & ATGTTCTC & ACCTCTTC & AGGTCTTA\\ 
135 & GCCAGACT & TTTACGGC & GGTAGGTG & TAAGACCT & CGCCCTGA & CGTAGAAC\\ 
136 & TAGCCGCC & TTGGGGGA & CGGAATGA & CGGAGCCT & TCCGCCAC & CCCTGACA\\ 
137 & GAGCTCAC & CTTGTACT & CTTATCCT & TCACCAGG\\ 
138 & GCGGATGC & GAGGTTAG & AGTCTGGC & AACGGGTT\\ 
139 & TCCCAATC & TTTATACC & ATTCTGCT & ATAGACGG & AACTGCGC & GTCACTGC\\ 
140 & AGGGATTG & AGTACCCA & TGTGTCCG & AGATGGCA & TATGACAG & GTGAGCAC\\ 
141 & TTCCGCGC & ATAAAACT & TGGTCGCA & TGTCAGGG & GTCGGGTC & GCCGTAAA\\ 
142 & TCAGGGCG & AGGTCTTA & GGCGGCTA & CGTCTATT & TAACGTGC & CTGATCAC\\ 
143 & GGACTGTT & CTACATAG & TGACTTGG & ACGAGGTT\\ 
144 & TCAGCGGC & AACCCGTT & CAATCCCT & AGCCGTTC\\ 
145 & GCATGCCG & GTAGCACC & CGAACGAG & ATCGTCCT & AGACGACG & GATCTCCT\\ 
146 & ATGACTCC & AGGAGAGG & GTGAGCTC & CTACGTGG & TGACGAAG & AATACCGT\\ 
147 & ATAACCAT & CCTGGTGA & AGGATAAG & GCAGTGAC & TGTAATAT & TGGGTACT\\ 
148 & CTGTCATA & TGCCATCT & GATTGGGA & GAGTCCAG & GCAATAAA & GAAACTGC\\ 
149 & TTCATTAC & GTTCTACG & CGATGCTT & TTGCCACA\\ 
150 & CCTACGCC & GAACGGCT & GGAGTCAT & CTCGGCAT\\ 
151 & AGTTTCCT & TTGAGACT & ACCCGAGC & TCGTTCGA & GCCAAAAT & CAAGGACG\\ 
152 & AGTTAGAT & CCATGCGA & AACAGTCC & GTATAGCG & CCCAGTGA & GCTCGACA\\ 
153 & GCGTTAAA & AACCTCGT & CCAAGTCA & AGGAGATC & GACCACGT & CCTCTCCT\\ 
154 & CTTCGTCA & CCACGTAG & CGGCATGC & AGCACGCA & TATTTGAC & GTCTTGCG\\ 
155 & AAGTGTTA & ATGCCGAG & ATCTAACT & TGATTTTT\\ 
156 & CCCGACCT & GTGCTCAC & GCACGAGA & AGGCTTCC\\ 
157 & TCACTGGG & CGCTATAC & AGGAAACT & CGAGGGGC & GAAGATAG & AGAGCATA\\ 
158 & TTAAGATG & TGTGGCAA & AAGCATCG & CGTCCTTG & GCATCCGC & TCGCATGG\\ 
159 & AAGAATGG & TCTACATC & GTAATGAA & GGATAGAT & AAGCAGAG & CATTTTGA\\ 
160 & TTCAGCGG & GTGATCAG & GCACGTTA & AATAGACG & GCGCGGAA & GACCTCCG\\ 
161 & AAGAGCAA & ACGGTATT & ATCCCGTT & CACAGCGC\\ 
162 & ACATGAGG & AAAAATCA & CCATTCTT & GTGCCAGT\\ 
163 & TTACCCTC & GCAGTTTC & TTTATTGC & CTGGACTC & ATGGTTAT & CCCTTCCC\\ 
164 & TTACCAGC & GTGCGGCC & AGGTCGGG & ACAGCAGT & ACCCATGG & CGCACCAC\\ 
165 & TATTTCCT & GGAAGCCT & TCTCGTGC & CGGAGGTC & GCCGCTGA & GATGTAGA\\ 
166 & CTCTGCTT & ATCTATCC & CCGCTGAA & AGGATTAT & TCAATAAT & GCACCTGC\\ 
167 & TCCCTGTC & TGTCGAGC & CCAGCGAC & CACTTCTG\\ 
168 & TTCTAGCC & ACTGGCAC & GCTGGTAA & TCTGGGCT\\ 
169 & CAGCTTGA & CGCAAGAC & GTCAAATA & TGCGTGCT & TTTAACGC & TTTAGACC\\ 
170 & ATTCAGAA & TCGAGACC & TTGCTCTT & GGACGTCT & CCATCAGA & TGGAGCTG\\ 
171 & TGGACCAC & GCGCTGTG & AACGGGAT & GGGAAGGG & GGCGTAGG & GGCCGCAC\\ 
172 & CCATGGGT & ACTGCTGT & GAGGGTAA & CACGTTGG & ACTCCTAC & TGCCCGCC\\ 
173 & TTCAACAA & TCAAAATG & ATATTGTA & GCGAGATA\\ 
174 & CTACATTT & AGCCCAGA & TTCTGAAT & CGCGGGCT\\ 
175 & AACTACTG & TATGCTCT & CTATCTTC & GCCCCTCG & CATCTTAA & ATATAACT\\ 
176 & CTAGTAAC & CGACCGTT & GACAGGGA & CCACCGCA & TCTTTCTC & GTATATCA\\ 
177 & CTAGGACC & CAGAAGTG & GTCGCTGG & GGTCTAAA & TAACACTT & GGTCTCGA\\ 
178 & TCTGATGG & AGACGTCC & TCAAGCTG & CTGCCCAG & TGATTAGC & GCTTAGGG\\ 
179 & CATTCTAT & AGCCCGCG & GTTACTAG & GCGCCTTG\\ 
180 & GATGATCT & GTGGTGCG & CACAATTT & GCTGGAAG\\ 
181 & GAGAAAGA & TGCGGTGG & CAGTAGTT & GTGATTAG & AATTCGCT & AAACCGGA\\ 
182 & TATCCGCC & TATCTCGC & TACAATAT & AGTTATAT & CCTCATGT & AACGGTCG\\ 
183 & ATTAAGCT & GTGGAATG & TTGTTGAA & CATACTCC & TCATTAGG & GTCCACCG\\ 
184 & TTCCTAAC & GCAGGTGC & ATTATTGA & ATAATCCT & AGGAAATA & CCACTGCC\\ 
185 & AGGGTTCA & CAGCTCCA & AAGCCTGT & CGAGGCGT\\ 
186 & GACTCTCT & CAAGGCGC & AGCTTAAT & TAGTAAGG\\ 
187 & TGTTTTAA & GGCGGGCA & GTAGGAGT & CCAACGTG & GTGGTCCA & CGGCGACG\\ 
188 & CTTTTCTG & CTTCCAGC & AAATTGTG & GGCAGTGG & GGCTAGAA & CATTCCAC\\ 
189 & CTTGGGGG & TGATATAC & CTCACTAA & ACTCCTTC\\ 
190 & GTCCTATC & CCTTACTA & GCGGTACA & CACCAAGA\\ 
191 & AACCTACC & CCCTAAGC & GCTAATCA & CTGGGCAG & GGTCCTAG & GTTGCAGA\\ 
192 & TTACAGCG & ACGCCTCG & ACAGGCTT & CGTCGCCG & AAATGTAG & CCATGAAT\\ 
193 & TGTACCGC & ATTCATGG & AGATCATC & CATTGACG & TCGTTACG & TTTTTTTT\\ 
194 & CCTAATGA & GGAGTATG & GTTAGGAA & TACCTGCT & CCTGAATG & TTTTTTTT\\ 
195 & CTCCTACG & CTCAATAG & TGAACCCT & TACTATCG\\ 
196 & AGAGTTCT & ACCAGCCA & AGAGAGTC & GCAGCCCT\\ 
197 & CGTAACGA & CGTCAATG & TTAAAACA & ACAACCGA\\ 
198 & CATTCAGG & AGCAGGTA & CAGAAAAG & GAACCGAC\\ 
199 & TTTTTTTT & CGATAGTA & GGTAGGTT & TTTTTTTT\\ 
200 & TTTTTTTT & AGGGCTGC & CCCCCAAG & TTTTTTTT\\ 
201 & TTTTTTTT & CGGTGGAC & AGAACTCT & TTTTTTTT\\ 
202 & TTTTTTTT & TCTTGGTG & CGTAGGAG & TTTTTTTT\\ 
203 & TTTTTTTT & TCGGTTGT & CGCTGTAA & TTTTTTTT\\ 
204 & TTTTTTTT & GTCGGTTC & GATAGGAC & TTTTTTTT\\ 
205 & TTTTTTTT & TCCGGTTT & AGCGAATT & CTAATCAC & GGCGGATA & TGGCTGGT\\ 
206 & TTTTTTTT & GAAGGAGT & TTAGTGAG & TCTGCAAC & ATAGAATG & CTATTGAG\\ 
207 & CAGACTCT & TTTTTTTT & TTTTTTTT & GAGCCGAT\\ 
208 & CAGGGTAC & TTTTTTTT & TTTTTTTT & GGACGAGC\\ 
209 & TCTGATAT & TTTTTTTT & TTTTTTTT & GCACTGCC\\ 
210 & CGCTGATC & TTTTTTTT & TTTTTTTT & GGGGATGA\\ 
211 & CAAGGGAT & ATCGGCTC & CTTCGGAT & TCTCTGGC\\ 
212 & TACATCTC & GCTCGTCC & TCACCATA & ATCCACCA\\ 
213 & AGAGTAGG & GGCAGTGC & TCAGCCAA & ACAGTGGG\\ 
214 & GGAAGGAT & TCATCCCC & TTTAAGAA & TGATCGCA\\ 
215 & TATCATTG & GCCAGAGA & TGCTAACT & GTATGACG\\ 
216 & TGCATGGG & TGGTGGAT & CGCGAAGC & TGTCGTGA\\ 
217 & AGATGTTC & CCCACTGT & TTATATTA & CAAAAGAT\\ 
218 & TAACCTGG & TGCGATCA & ATCCCGGA & CACGTAAA\\ 
219 & ACCTGGTC & CGTCATAC & TAAAAGGT & AGAAGTAA\\ 
220 & CCGATCAT & TCACGACA & AATCTCCC & GCTAGATG\\ 
221 & ACAATGGT & ATCTTTTG & AAGCGCAC & GTAAACCG\\ 
222 & CCGCAACC & TTTACGTG & TTTAGGGA & ATACAAGA\\ 
223 & TACCCTAC & TTACTTCT & AGTGTAAG & CAACCCAT\\ 
224 & CCAGACCG & CATCTAGC & ATTACCTT & CACCGCTC\\ 
225 & CCGAATTA & CGGTTTAC & CGAGATGT & TTACTTGC\\ 
226 & CCACAAAG & TCTTGTAT & AAAGACTG & TGCGAGAT\\ 
227 & AAAGTAAT & ATGGGTTG & CACTAAAC & TATAGCAA\\ 
228 & GTGCTCTT & GAGCGGTG & TTAAGAGC & CTCGCTTC\\ 
229 & TCCCAAGT & GCAAGTAA & AAAAGGAA & TTATGAGA\\ 
230 & GCGTAAAG & ATCTCGCA & TTCTGGTG & CCAAAACT\\ 
231 & GAACATAA & TTGCTATA & TCGGACTG & TGGGCGTA\\ 
232 & ACGGACTT & GAAGCGAG & GAAGTTTT & GCGACTCA\\ 
233 & ATAATAGT & TCTCATAA & CAGGCAAG & GGTGTCAG\\ 
234 & GTATGATT & AGTTTTGG & CTGCTGAA & TGTCCGTG\\ 
235 & GGTGTAGG & TACGCCCA & TAAACTAC & TTTTCCAC\\ 
236 & CTCAGCAA & TGAGTCGC & TTGAGTCG & GTGGATAA\\ 
237 & ATATAGGT & CTGACACC & CAGCTCCC & AGGTAAGT\\ 
238 & AGCAATAT & CACGGACA & GGAACAAC & GGGTGTCA\\ 
239 & AGGGCAAC & GTGGAAAA & TTGTGATC & CCCTCCCA\\ 
240 & CGTGGATC & TTATCCAC & GCCTCAGT & CCATTTCA\\ 
241 & GATAATAC & ACTTACCT & CCACATGA & AGGTCTAC\\ 
242 & CGACCATT & TGACACCC & GAGTCGGC & CTGACAGA\\ 
243 & ACAAAGTC & TGGGAGGG & CCTCAACA & TCTTGAAT\\ 
244 & GCGTATGC & TGAAATGG & AACCCTCC & TACGTTTT\\ 
245 & AGTTTCTG & GTAGACCT & CGTAGGCT & AGAGCGGT\\ 
246 & AAAAGATG & TCTGTCAG & CCCAAGCT & TAGCATAA\\ 
247 & AATAGGCT & ATTCAAGA & GCTGGACG & TGCGCGAG\\ 
248 & ACCATCGC & AAAACGTA & AAACTAAA & TGACGTTT\\ 
249 & CGCACGGC & ACCGCTCT & CTCTCCGC & CGGCGAAA\\ 
250 & CAACAGTT & TTATGCTA & ACTTGGAC & CGGGCATC\\ 
251 & ATACGTGC & CTCGCGCA & ACGCTCAG & CCGGTCTG\\ 
252 & ATCACCTC & AAACGTCA & TAGCCCTA & CGGATCAA\\ 
253 & TATTAAGT & TTTCGCCG & TGAGCGAT & CAATGACC\\ 
254 & TCGTATTT & GATGCCCG & TGTGTATG & CTCACTTT\\ 
255 & TCACCCTA & CAGACCGG & TCAGGATC & ATTTTGGA\\ 
256 & ATTCCCAG & TTGATCCG & CGCATTAC & AACGTTAT\\ 
257 & CCATATGA & GGTCATTG & TGGATATC & TCCGTGCA\\ 
258 & GCAGGGTG & AAAGTGAG & CCAGCCGC & GCGCAGGG\\ 
259 & TGATATGG & TCCAAAAT & GTGCAACG & GGAAACAC\\ 
260 & GAAAGATC & ATAACGTT & GGGATTTG & TTCCGCCT\\ 
261 & TCGTTTAC & TGCACGGA & ACGAGAGT & GCTGGCAT\\ 
262 & CCGTCGAG & CCCTGCGC & GACAAACG & CATATCCT\\ 
263 & CTCGAATT & GTGTTTCC & ATGCTAGC & GCCACACG\\ 
264 & TTACCTAT & AGGCGGAA & AAATTGAC & ATTTCTAC\\ 
265 & ATCGACTA & ATGCCAGC & ATTACTGG & CATTAAGA\\ 
266 & AAGTAAAC & AGGATATG & GCGCCAGA & CCACAATG\\ 
267 & TAAGTAGT & CGTGTGGC & TGCAAGCC & CTCACCTG\\ 
268 & TCACGATT & GTAGAAAT & TTTACAAA & GACGGACA\\ 
269 & CTTCTACC & TCTTAATG & TATAATTC & ATCCTGTG\\ 
270 & CCGTCACG & CATTGTGG & TTCGCCGC & GTCGTTTT\\ 
271 & TAAGAGGT & CAGGTGAG & CGGGCCCA & ATTGTACC\\ 
272 & GATTCTAA & TGTCCGTC & AATGAGCT & TGGGATCT\\ 
273 & CCGTAATC & CACAGGAT & TCTTGCTT & ATCTTAGT\\ 
274 & GCCGATGG & AAAACGAC & GCGGGCCG & GGAAGGAC\\ 
275 & TGTTTACT & GGTACAAT & TCTTACAC & GGGTAGGA\\ 
276 & CGACCTCT & AGATCCCA & ATGTTAGG & AGCTAACA\\ 
277 & ACTTGGGG & ACTAAGAT & CGTTGCGA & ATTTTCGC\\ 
278 & GCCTAAAG & GTCCTTCC & ACGGTAAT & CGGGGGTG\\ 
279 & ACGGAAAC & TCCTACCC & GCGCGGTT & GATGTGCC\\ 
280 & GAAATATC & TGTTAGCT & TTTGAAGA & AGGCTCCG\\ 
281 & CTAATATA & GCGAAAAT & CTTCCATG & TAATTATA\\ 
282 & AGTTCGTA & CACCCCCG & AAACTGAG & GAGAACAT\\ 
283 & TTTTCGCA & GGCACATC & GAGCCTAC & CCGTCTAT\\ 
284 & CGGAATGA & CGGAGCCT & TCCGCCAC & CCCTGACA\\ 
285 & GTTACCGC & TATAATTA & AGAACATT & CGGGTTCG\\ 
286 & ACAGGACG & ATGTTCTC & ACCTCTTC & AGGTCTTA\\ 
287 & ATTCTGCT & ATAGACGG & AACTGCGC & GTCACTGC\\ 
288 & TGGTCGCA & TGTCAGGG & GTCGGGTC & GCCGTAAA\\ 
289 & CAATTCCC & CGAACCCG & CGAATCGC & AGGACGAT\\ 
290 & CCTAGGCG & TAAGACCT & GCCGGAGT & AGACGTGC\\ 
291 & AGGATAAG & GCAGTGAC & TGTAATAT & TGGGTACT\\ 
292 & GCCAGACT & TTTACGGC & GGTAGGTG & TAAGACCT\\ 
293 & CGAACGAG & ATCGTCCT & AGACGACG & GATCTCCT\\ 
294 & CTCGCGTG & GCACGTCT & ACGACTCG & TCGAACGA\\ 
295 & AGGGATTG & AGTACCCA & TGTGTCCG & AGATGGCA\\ 
296 & TCAGGGCG & AGGTCTTA & GGCGGCTA & CGTCTATT\\ 
297 & CCAAGTCA & AGGAGATC & GACCACGT & CCTCTCCT\\ 
298 & ACCCGAGC & TCGTTCGA & GCCAAAAT & CAAGGACG\\ 
299 & CTGTCATA & TGCCATCT & GATTGGGA & GAGTCCAG\\ 
300 & GCACGTTA & AATAGACG & GCGCGGAA & GACCTCCG\\ 
301 & ATGACTCC & AGGAGAGG & GTGAGCTC & CTACGTGG\\ 
302 & AAGCATCG & CGTCCTTG & GCATCCGC & TCGCATGG\\ 
303 & TTTATTGC & CTGGACTC & ATGGTTAT & CCCTTCCC\\ 
304 & TCTCGTGC & CGGAGGTC & GCCGCTGA & GATGTAGA\\ 
305 & CTTCGTCA & CCACGTAG & CGGCATGC & AGCACGCA\\ 
306 & AGTTAGAT & CCATGCGA & AACAGTCC & GTATAGCG\\ 
307 & AACGGGAT & GGGAAGGG & GGCGTAGG & GGCCGCAC\\ 
308 & AAGAATGG & TCTACATC & GTAATGAA & GGATAGAT\\ 
309 & GTCAAATA & TGCGTGCT & TTTAACGC & TTTAGACC\\ 
310 & TCACTGGG & CGCTATAC & AGGAAACT & CGAGGGGC\\ 
311 & TTACCAGC & GTGCGGCC & AGGTCGGG & ACAGCAGT\\ 
312 & CTCTGCTT & ATCTATCC & CCGCTGAA & AGGATTAT\\ 
313 & GTCGCTGG & GGTCTAAA & TAACACTT & GGTCTCGA\\ 
314 & CTATCTTC & GCCCCTCG & CATCTTAA & ATATAACT\\ 
315 & CCATGGGT & ACTGCTGT & GAGGGTAA & CACGTTGG\\ 
316 & ATTATTGA & ATAATCCT & AGGAAATA & CCACTGCC\\ 
317 & ATTCAGAA & TCGAGACC & TTGCTCTT & GGACGTCT\\ 
318 & TACAATAT & AGTTATAT & CCTCATGT & AACGGTCG\\ 
319 & GTAGGAGT & CCAACGTG & GTGGTCCA & CGGCGACG\\ 
320 & AAATTGTG & GGCAGTGG & GGCTAGAA & CATTCCAC\\ 
321 & TCTGATGG & AGACGTCC & TCAAGCTG & CTGCCCAG\\ 
322 & CTAGTAAC & CGACCGTT & GACAGGGA & CCACCGCA\\ 
323 & ACAGGCTT & CGTCGCCG & AAATGTAG & CCATGAAT\\ 
324 & ATTAAGCT & GTGGAATG & TTGTTGAA & CATACTCC\\ 
325 & GCTAATCA & CTGGGCAG & GGTCCTAG & GTTGCAGA\\ 
326 & GAGAAAGA & TGCGGTGG & CAGTAGTT & GTGATTAG\\ 
327 & TGTACCGC & ATTCATGG & AGATCATC & CATTGACG\\ 
328 & CCTAATGA & GGAGTATG & GTTAGGAA & TACCTGCT\\ 
329 & TTAGTGAG & TCTGCAAC & ATAGAATG & CTATTGAG\\ 
330 & AGCGAATT & CTAATCAC & GGCGGATA & TGGCTGGT\\ 
331 & ACCTTTTA & ATCATCTC\\ 
332 & GGGAGATT & TAGCGGTC\\ 
333 & CCTACTCT & CACCTTTC\\ 
334 & ATCCTTCC & AACATCTC\\ 
335 & CTTACACT & TTCGGGAG\\ 
336 & CCCATGCA & GAAAGATA\\ 
337 & ACATCTCG & GAAAGGTG\\ 
338 & CCAGGTTA & AGTGGCTC\\ 
339 & GACCAGGT & CTCCTAGG\\ 
340 & ATGATCGG & TCATATGT\\ 
341 & TTCCTTTT & CAACCAAA\\ 
342 & CACCAGAA & GAGCCACT\\ 
343 & GTAGGGTA & CCGCTGGG\\ 
344 & AAAACTTC & ACATATGA\\ 
345 & TAATTCGG & AGCCCGGC\\ 
346 & TTCAGCAG & CCTTACTC\\ 
347 & GTAGTTTA & CCCAGCGG\\ 
348 & CGACTCAA & CCTCACAG\\ 
349 & ACTTGGGA & TTAGGATC\\ 
350 & CTTTACGC & GCTTGAAC\\ 
351 & GATCACAA & ATATGGCA\\ 
352 & AAGTCCGT & GTCGGTAG\\ 
353 & TCATGTGG & GATCCTAA\\ 
354 & AATCATAC & TCACGGTT\\ 
355 & CCTACACC & ACAGATGC\\ 
356 & TTGCTGAG & TTCGACGC\\ 
357 & AGCCTACG & ATAGCCAG\\ 
358 & AGCTTGGG & AACCGTGA\\ 
359 & GTTGCCCT & CGACCGAT\\ 
360 & TTTAGTTT & GCGTCGAA\\ 
361 & GTATTATC & GCGTCAAT\\ 
362 & GTCCAAGT & AAGGGCAA\\ 
363 & CTGAGCGT & ATCGGTCG\\ 
364 & TAGGGCTA & AAATCTGA\\ 
365 & CAGAAACT & ATGTACCG\\ 
366 & CATCTTTT & GGGGTCGT\\ 
367 & GATCCTGA & GAGATTAC\\ 
368 & GCGATGGT & GATGGGAA\\ 
369 & GATATCCA & CGGTACAT\\ 
370 & AACTGTTG & TGTCCTAT\\ 
371 & GCACGTAT & GTCTTGAG\\ 
372 & GAGGTGAT & ACGTAAAG\\ 
373 & ACTCTCGT & TCAGTCGT\\ 
374 & CGTTTGTC & ATAGGACA\\ 
375 & TAGGGTGA & CGTATTAG\\ 
376 & GTCAATTT & CTTTACGT\\ 
377 & TCATATGG & ACACACCA\\ 
378 & TCTGGCGC & CGGCTTAG\\ 
379 & GGCTTGCA & CTAATACG\\ 
380 & TTTGTAAA & CTTACCTA\\ 
381 & GTAAACGA & TAGGTTTC\\ 
382 & CTCGACGG & TTAATTTG\\ 
383 & TGGGCCCG & GGGTTTGT\\ 
384 & ATAGGTAA & CTGGGTTT\\ 
385 & AAGCAAGA & GAAACCTA\\ 
386 & GTTTACTT & AGGGATGG\\ 
387 & ACTACTTA & AGACTTAG\\ 
388 & AATCGTGA & CGGTTTCT\\ 
389 & TCGCAACG & GAAGCGTA\\ 
390 & ATTACCGT & CCATCCCT\\ 
391 & ACCTCTTA & GCACAATA\\ 
392 & TCTTCAAA & AGAAACCG\\ 
393 & GATTACGG & CGGCTTAG\\ 
394 & CTCAGTTT & AAGATTCA\\ 
395 & GTAGGCTC & TATTGTGC\\ 
396 & GTGGCGGA & CCTAAGGT\\ 
397 & CCCCAAGT & GCTACATG\\ 
398 & CTTTAGGC & CCAAGGAG\\ 
399 & GCGCAGTT & TCCTTGAT\\ 
400 & GATATTTC & AAGGACAC\\ 
401 & GCGATTCG & CATGTAGC\\ 
402 & TACGAACT & CAACGCAC\\ 
403 & TGCGAAAA & GCAAGCCG\\ 
404 & TCATTCCG & TCCCCCAA\\ 
405 & CGTCGTCT & TGGGTAAG\\ 
406 & CGAGTCGT & GTGCGTTG\\ 
407 & AGCAGAAT & GGTATAAA\\ 
408 & TAGCCGCC & TTGGGGGA\\ 
409 & GGGAATTG & AGTACAAG\\ 
410 & ATTTTGGC & GGAACGGG\\ 
411 & TCCCAATC & TTTATACC\\ 
412 & TTCCGCGC & ATAAAACT\\ 
413 & CTCGTTCG & GGTGCTAC\\ 
414 & CACGCGAG & CTATGTAG\\ 
415 & ATAACCAT & CCTGGTGA\\ 
416 & CGCCCTGA & CGTAGAAC\\ 
417 & GCATGCCG & GTAGCACC\\ 
418 & GCTCGGGT & AGTCTCAA\\ 
419 & TATGACAG & GTGAGCAC\\ 
420 & TAACGTGC & CTGATCAC\\ 
421 & GCGTTAAA & AACCTCGT\\ 
422 & AGTTTCCT & TTGAGACT\\ 
423 & GCAATAAA & GAAACTGC\\ 
424 & TTCAGCGG & GTGATCAG\\ 
425 & TGACGAAG & AATACCGT\\ 
426 & TTAAGATG & TGTGGCAA\\ 
427 & TTACCCTC & GCAGTTTC\\ 
428 & TATTTCCT & GGAAGCCT\\ 
429 & TATTTGAC & GTCTTGCG\\ 
430 & CCCAGTGA & GCTCGACA\\ 
431 & TGGACCAC & GCGCTGTG\\ 
432 & AAGCAGAG & CATTTTGA\\ 
433 & CAGCTTGA & CGCAAGAC\\ 
434 & GAAGATAG & AGAGCATA\\ 
435 & ACCCATGG & CGCACCAC\\ 
436 & TCAATAAT & GCACCTGC\\ 
437 & CTAGGACC & CAGAAGTG\\ 
438 & AACTACTG & TATGCTCT\\ 
439 & ACTCCTAC & TGCCCGCC\\ 
440 & TTCCTAAC & GCAGGTGC\\ 
441 & CCATCAGA & TGGAGCTG\\ 
442 & TATCCGCC & TATCTCGC\\ 
443 & TGTTTTAA & GGCGGGCA\\ 
444 & CTTTTCTG & CTTCCAGC\\ 
445 & TGATTAGC & GCTTAGGG\\ 
446 & TCTTTCTC & GTATATCA\\ 
447 & TTACAGCG & ACGCCTCG\\ 
448 & TCATTAGG & GTCCACCG\\ 
449 & AACCTACC & CCCTAAGC\\ 
450 & AATTCGCT & AAACCGGA\\ 
451 & TCGTTACG & TTTTTTTT\\ 
452 & CCTGAATG & TTTTTTTT\\ 
453 & TTTTTTTT & GAAGGAGT\\ 
454 & TTTTTTTT & TCCGGTTT
\end{xtabular}
\end{center}

\vskip6ex
\noindent The following bricks were used in all structures investigated:\\
\noindent{\scriptsize{}1, 2, 3, 4, 5, 6, 9, 10, 11, 12, 13, 14, 17, 18, 21, 22, 29, 30, 35, 36, 41, 42, 45, 46, 53, 54, 59, 60, 65, 66, 71, 72, 77, 78, 83, 84, 89, 90, 93, 94, 101, 102, 107, 108, 113, 114, 119, 120, 125, 126, 131, 132, 137, 138, 143, 144, 149, 150, 155, 156, 161, 162, 167, 168, 173, 174, 179, 180, 185, 186, 189, 190, 195, 196, 197, 198, 199, 200, 201, 202, 203, 204}

\vskip1ex
\noindent In addition to the bricks common to all structures, the following bricks were used for each class of structure studied.

\vskip1ex\noindent The \textbf{\color{colourNo}no-BB} system (330 bricks in total):\\
\noindent{\scriptsize{}207, 208, 209, 210, 211, 212, 213, 214, 215, 216, 217, 218, 219, 220, 221, 222, 223, 224, 225, 226, 227, 228, 229, 230, 231, 232, 233, 234, 235, 236, 237, 238, 239, 240, 241, 242, 243, 244, 245, 246, 247, 248, 249, 250, 251, 252, 253, 254, 255, 256, 257, 258, 259, 260, 261, 262, 263, 264, 265, 266, 267, 268, 269, 270, 271, 272, 273, 274, 275, 276, 277, 278, 279, 280, 281, 282, 283, 284, 285, 286, 287, 288, 289, 290, 291, 292, 293, 294, 295, 296, 297, 298, 299, 300, 301, 302, 303, 304, 305, 306, 307, 308, 309, 310, 311, 312, 313, 314, 315, 316, 317, 318, 319, 320, 321, 322, 323, 324, 325, 326, 327, 328, 329, 330, 331, 332, 333, 334, 335, 336, 337, 338, 339, 340, 341, 342, 343, 344, 345, 346, 347, 348, 349, 350, 351, 352, 353, 354, 355, 356, 357, 358, 359, 360, 361, 362, 363, 364, 365, 366, 367, 368, 369, 370, 371, 372, 373, 374, 375, 376, 377, 378, 379, 380, 381, 382, 383, 384, 385, 386, 387, 388, 389, 390, 391, 392, 393, 394, 395, 396, 397, 398, 399, 400, 401, 402, 403, 404, 405, 406, 407, 408, 409, 410, 411, 412, 413, 414, 415, 416, 417, 418, 419, 420, 421, 422, 423, 424, 425, 426, 427, 428, 429, 430, 431, 432, 433, 434, 435, 436, 437, 438, 439, 440, 441, 442, 443, 444, 445, 446, 447, 448, 449, 450, 451, 452, 453, 454}

\vskip1ex\noindent The \textbf{\color{colourEdge}edge-BB} system (268 bricks in total):\\
\noindent{\scriptsize{}8, 15, 19, 23, 26, 28, 31, 34, 37, 40, 43, 47, 50, 52, 55, 58, 61, 64, 67, 70, 73, 76, 79, 82, 85, 88, 91, 95, 98, 100, 103, 106, 109, 112, 115, 118, 121, 124, 127, 130, 133, 136, 139, 142, 145, 148, 151, 154, 157, 160, 163, 166, 169, 172, 175, 178, 181, 184, 187, 191, 194, 205, 208, 210, 211, 212, 215, 217, 221, 222, 224, 226, 227, 228, 231, 233, 237, 238, 240, 242, 243, 244, 247, 249, 253, 254, 256, 258, 259, 260, 263, 265, 269, 270, 272, 274, 275, 276, 279, 281, 285, 286, 288, 290, 291, 292, 295, 297, 301, 302, 304, 306, 307, 308, 311, 313, 317, 318, 320, 322, 323, 324, 327, 329, 332, 334, 335, 336, 339, 341, 345, 346, 348, 350, 351, 352, 355, 357, 361, 362, 364, 366, 367, 368, 371, 373, 377, 378, 380, 382, 383, 384, 387, 389, 393, 394, 396, 398, 399, 400, 403, 405, 409, 410, 412, 414, 415, 416, 419, 421, 425, 426, 428, 430, 431, 432, 435, 437, 441, 442, 444, 446, 447, 448, 451, 453}

\vskip1ex\noindent The \textbf{\color{colourFace}face-BB} system (268 bricks in total):\\
\noindent{\scriptsize{}7, 16, 20, 24, 25, 27, 32, 33, 38, 39, 44, 48, 49, 51, 56, 57, 62, 63, 68, 69, 74, 75, 80, 81, 86, 87, 92, 96, 97, 99, 104, 105, 110, 111, 116, 117, 122, 123, 128, 129, 134, 135, 140, 141, 146, 147, 152, 153, 158, 159, 164, 165, 170, 171, 176, 177, 182, 183, 188, 192, 193, 206, 207, 209, 213, 214, 216, 218, 219, 220, 223, 225, 229, 230, 232, 234, 235, 236, 239, 241, 245, 246, 248, 250, 251, 252, 255, 257, 261, 262, 264, 266, 267, 268, 271, 273, 277, 278, 280, 282, 283, 284, 287, 289, 293, 294, 296, 298, 299, 300, 303, 305, 309, 310, 312, 314, 315, 316, 319, 321, 325, 326, 328, 330, 331, 333, 337, 338, 340, 342, 343, 344, 347, 349, 353, 354, 356, 358, 359, 360, 363, 365, 369, 370, 372, 374, 375, 376, 379, 381, 385, 386, 388, 390, 391, 392, 395, 397, 401, 402, 404, 406, 407, 408, 411, 413, 417, 418, 420, 422, 423, 424, 427, 429, 433, 434, 436, 438, 439, 440, 443, 445, 449, 450, 452, 454}

\newpage
\vskip1ex\noindent The \textbf{\color{colourHalfFace}half-face-BB} system (299 bricks in total):\\
\noindent{\scriptsize{}7, 16, 20, 32, 38, 39, 44, 56, 62, 63, 69, 80, 86, 87, 92, 104, 110, 111, 117, 128, 134, 135, 141, 152, 158, 159, 165, 176, 182, 183, 188, 207, 209, 211, 213, 214, 215, 216, 217, 218, 219, 220, 221, 223, 225, 227, 229, 230, 231, 232, 233, 234, 235, 236, 237, 239, 241, 243, 245, 246, 247, 248, 249, 250, 251, 252, 253, 255, 257, 259, 261, 262, 263, 264, 265, 266, 267, 268, 269, 271, 273, 275, 277, 278, 279, 280, 281, 282, 283, 284, 285, 287, 289, 291, 293, 294, 295, 296, 297, 298, 299, 300, 301, 303, 305, 307, 309, 310, 311, 312, 313, 314, 315, 316, 317, 319, 321, 323, 325, 326, 327, 328, 329, 330, 331, 333, 335, 337, 338, 339, 340, 341, 342, 343, 344, 345, 347, 349, 351, 353, 354, 355, 356, 357, 358, 359, 360, 361, 363, 365, 367, 369, 370, 371, 372, 373, 374, 375, 376, 377, 379, 381, 383, 385, 386, 387, 388, 389, 390, 391, 392, 393, 395, 397, 399, 401, 402, 403, 404, 405, 406, 407, 408, 409, 411, 413, 415, 417, 418, 419, 420, 421, 422, 423, 424, 425, 427, 429, 431, 433, 434, 435, 436, 437, 438, 439, 440, 441, 443, 445, 447, 449, 450, 451, 452, 453, 454}

\vskip1ex\noindent The \textbf{\color{colourAll}all-BB} system (206 bricks in total):\\
\noindent{\scriptsize{}7, 8, 15, 16, 19, 20, 23, 24, 25, 26, 27, 28, 31, 32, 33, 34, 37, 38, 39, 40, 43, 44, 47, 48, 49, 50, 51, 52, 55, 56, 57, 58, 61, 62, 63, 64, 67, 68, 69, 70, 73, 74, 75, 76, 79, 80, 81, 82, 85, 86, 87, 88, 91, 92, 95, 96, 97, 98, 99, 100, 103, 104, 105, 106, 109, 110, 111, 112, 115, 116, 117, 118, 121, 122, 123, 124, 127, 128, 129, 130, 133, 134, 135, 136, 139, 140, 141, 142, 145, 146, 147, 148, 151, 152, 153, 154, 157, 158, 159, 160, 163, 164, 165, 166, 169, 170, 171, 172, 175, 176, 177, 178, 181, 182, 183, 184, 187, 188, 191, 192, 193, 194, 205, 206}

\end{document}